\documentclass[paper,a4paper,11pt,twoside]{JHEP3}
\usepackage{macros} 
\usepackage{amssymb,amsmath}
\usepackage{booktabs}
\usepackage{epsfig}
\makeatletter
\def\@xcmidrule{\ifx\@tempa\cmidrule\vskip-\@thisrulewidth
     \global\@lastruleclass=\@ne\else
     \ifx\@tempa\morecmidrules\vskip \cmidrulesep
     \global\@lastruleclass=\@ne\else
     \vskip \belowrulesep\global\@lastruleclass=\z@\fi\fi
     \ifnum0=`{\fi}}
\makeatother
\title{%
Heavy-strange meson decay constants 
in the continuum limit of quenched QCD
}

\author{%
\begin{flushleft}
\vspace{-0.5cm}
\vbox{
\epsfxsize=2.5 true cm
\epsfbox{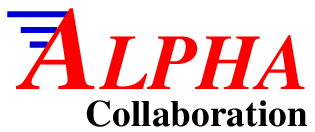}}
\end{flushleft}
}

\author{%
Michele Della Morte$^{\,a}$,
Stephan D\"urr$^{\,b}$,
Damiano Guazzini$^{\,c}$,
Jochen Heitger$^{\,d}$,
Andreas J\"uttner$^{\,e,f}$ and
Rainer Sommer$^{\,c}$\\
$^{\small a}$
CERN, Physics Department, TH Unit,\\\hspace{0.5em} 
CH-1211 Geneva~23, Switzerland\\
$^{\small b}$
Universit\"at Bern, Institut f\"ur Theoretische Physik,\\\hspace{0.5em} 
Sidlerstra{\ss}e~5, CH-3012 Bern, Switzerland\\
$^{\small c}$
Deutsches Elektronen-Synchrotron DESY, Zeuthen,\\\hspace{0.5em} 
Platanenallee~6, D-15738 Zeuthen, Germany\\
$^{\small d}$
Westf\"alische Wilhelms-Universit\"at M\"unster,
Institut f\"ur Theoretische Physik,\\\hspace{0.5em} 
Wilhelm-Klemm-Stra{\ss}e~9, D-48149 M\"unster, Germany\\
$^{\small e}$
University of Southampton, Department of Physics and Astronomy,\\\hspace{0.5em}
Highfield, Southampton~SO17~1BJ, United Kingdom\\
$^{\small f}$
Johannes Gutenberg-Universit\"at Mainz, Institut f\"ur Kernphysik,\\\hspace{0.5em}
Johann-Joachim-Becher-Weg~45, D-55099 Mainz, Germany\\\hspace{0.5em}
E-mail: \email{dellamor@mail.cern.ch,damiano.guazzini@desy.de,\\\hspace{0.5em}
heitger@uni-muenster.de,juettner@kph.uni-mainz.de,rainer.sommer@desy.de}
}

\preprint{%
CERN-PH-TH/2007-141\\
DESY 07-150\\
MS-TP-07-24\\
SHEP-07-32\\
SFB/CPP-07-52\\
\today
}

\abstract{%
We improve a previous quenched result for heavy-light pseudoscalar meson 
decay constants with the light quark taken to be the strange quark. 
A finer lattice resolution ($a\approx 0.05\,\Fm$) in the continuum limit 
extrapolation of the data computed in the static approximation is included. 
We also give further details concerning the techniques used in order to 
keep the statistical and systematic errors at large lattice sizes $L/a$ 
under control. 
Our final result, obtained by combining these data with determinations of 
the decay constant for pseudoscalar mesons around the ${\rm D_s}$, follows 
nicely the qualitative expectation of the $1/m$--expansion with a (relative) 
$1/m$--term of about $-0.5\,\GeV/\mps$. 
At the physical b-quark mass we obtain $F_{\rm B_{\rm s}}=193(7)$ MeV, where 
all errors apart from the quenched approximation are included. 
}


\begin{document}
\section{Introduction}
\label{Sec_intro}
A big experimental progress is expected for the next years in flavour 
physics, mainly due to the new LHCb experiment~\cite{Barsuk:2005ac} at CERN,  
CDF at Tevatron and the future super-B factories~\cite{Akeroyd:2004mj}.
It will then be possible to determine with a precision of a few percent all 
entries of the CKM matrix, which describes flavour changing currents in the 
Standard Model.
For this programme to be successful and to provide constraints on New Physics 
beyond the Standard Model, accurate theoretical predictions to be compared 
with the experimental results are extremely important.

Lattice QCD is the most appropriate tool for such computations, as they 
involve matrix elements of the operators in the Weak Effective Hamiltonian 
among hadronic states and they therefore require a non-perturbative approach. 

Still, b-quarks on the lattice pose a two-scale problem: 
the lattice spacing $a$ must be smaller than $1/m_{\rm b}$ and the size $L$ 
must be large enough such that the physics is not distorted by finite-size 
effects. 
Heavy Quark Effective Theory (HQET) on the lattice, as formulated by Eichten 
and Hill in~\Refs{stat:eichten,stat:eichhill1}, allows to circumvent the 
problem in a theoretically sound way. 
Formally, it consists in an expansion of the QCD action and correlation 
functions in inverse powers of the heavy quark mass.

Numerical applications have been plagued for a long time by the exponential 
growth of the noise-to-signal ratio in static-light correlation functions. 
This is due to the appearance of power divergences in the effective theory. 
Such divergences are non-universal, and we have given in~\cite{fbstat:pap1}
first evidences how the problem can be overcome by minimally modifying the 
Eichten-Hill discretization of the static action.

The result has been substantiated by successive 
studies~\cite{HQET:statprec,HQET:mb1m,ospin:Nf0,bb:pap1} and 
also in the theory with $N_{\rm f}=2$ dynamical quarks~\cite{zastat:Nf2}.
Most of the cases dealt with the non-perturbative renormalization of 
static-light operators in a finite-volume scheme.

Here we show that also for physically large volumes fine lattice resolutions 
can be reached and precise results obtained. 
To emphasize the importance of such studies, we find that the result for 
$F_{\rm B_{\rm s}}$ in the continuum limit changes by one standard deviation 
($7\%$) of the result quoted in~\Ref{fbstat:pap1}. 
At the same time, of course, we reduce the systematic uncertainty owing to
the extrapolation to zero lattice spacing. 

The dependence of the decay constant on the heavy quark mass can very well 
be reconstructed by combining the continuum static result with continuum 
results at quark masses around the physical charm quark mass. 
The connection of the two different regimes is smooth, once the 
renormalization and matching of the static result is taken into account with 
sufficient precision.

The paper is organized as follows. 
In \Sect{Sec_hlhphys} we describe the numerical setup used in the static 
approximation. 
Section~\ref{Sec_ana} deals with the fitting procedure adopted to extract 
effective energies and matrix elements. 
Results in the static approximation are collected in \Sect{Sec_res}, while 
\Sect{Sec_Fbs} contains details about the simulations with relativistic 
quarks around the charm and our central results. 
Conclusions are drawn in \Sect{Sec_concl}. 
The discussion of contaminations from excited states in static-light 
correlation functions is deferred to the appendix.

\section{Heavy-light hadron physics in the static approximation}
\label{Sec_hlhphys}
We consider a heavy-light meson system in the framework of quenched O($a$)
improved lattice QCD with Schr\"odinger functional boundary conditions,
where the heavy quark flavour is treated at the leading order of HQET, 
the static approximation.
The basic setup of the computations presented here follows our earlier 
determinations of hadron properties by means of numerical simulations of the 
QCD Schr\"odinger functional in physically large volumes, see 
e.g.~\Refs{msbar:pap2,msbar:pap3,mcbar:RS02} for studies in the strange and 
charm quark sectors 
and~\Refs{fbstat:pap1,HQET:statprec,HQET:mb1m,zastat:Nf2} for B-physics
applications.

A particularly important technical ingredient of extracting B-meson masses
and matrix elements from lattice HQET is the use of the alternative 
discretizations of the static theory, which were introduced 
in~\Refs{fbstat:pap1,HQET:statprec} to temper the well-known problem of 
exponential degradation of the signal-to-noise ratio encountered in 
static-light correlation functions when computed with the traditional 
Eichten-Hill \cite{stat:eichhill1} lattice action for the static quark.

Among the static quark actions $S_{\rm h}$ studied in detail 
in~\Ref{HQET:statprec}, in the present work we restrict ourselves to the 
``HYP-action'', which turns out to yield the largest gain (of more than one 
order of magnitude compared to Eichten-Hill at time separations of about 
$1.5\,\Fm$) in the signal-to-noise ratios of static-light correlation 
functions.
This action is constructed from the Eichten-Hill action through replacing 
the temporal parallel transporters in the lattice derivative acting on the 
heavy quark field by the HYP-link that is obtained by a sensibly chosen 
smearing prescription for the gauge links located within the neighbouring 
hypercube \cite{HYP:HK01}.
Actually, in the context of static quarks, there are two favourable 
parameterizations of the HYP-link available (referred to as ``HYP1'' and 
``HYP2'' in the following), with the second being even superior to the 
first.
For more details the reader may consult~\Ref{HQET:statprec}.\footnote{
As the static quark action HYP2 became available only at a final stage of
this project, only the simulations at our finest lattice resolution 
($\beta=6.45$) were done for both HYP-actions, HYP1 and HYP2.
}
\subsection{%
Correlation functions and their quantum mechanical representation}
\label{Sec_hlhphys_CFs}
Our starting point are B-meson correlation functions defined in the 
Schr\"odinger functional with a vanishing background gauge 
field~\cite{zastat:pap1}.
Quarks with finite masses, also referred to as relativistic quarks, are 
labelled with an index ``l'', and the static ones with an index ``h''. 
The O($a$) improved axial vector current in static approximation is then
defined as
\be
\Astatimpr(x)= 
\Astat(x)+a\,\castat\delta\Astat(x)
\label{astatimpr}
\ee
with the (bare) unimproved current
\be
\Astat(x)=\lightb(x)\gamma_0\gamma_5\heavy(x)
\label{astat}
\ee
and its O($a$) counterterm
\be
\delta\Astat(x)=
\lightb(x)\gamma_j\gamma_5{1\over 2}
\left(\lnab{j}+\lnabstar{j}\right)\heavy(x)
\label{dastat}
\ee
multiplied by the improvement coefficient $\castat$.
In order to suppress contributions from excited B-meson states to the 
correlation functions of interest, we implement wave functions 
$\omega({\bf x})$ at the boundaries of the Schr\"odinger functional such 
that an interpolating B-meson field is constructed in terms of the boundary 
quark fields $\zetal$ and $\zetahb$.
In this way, the correlation function of the static axial current, 
$f_{\rm A}^{\rm stat}$, takes the form
\be
f_{\rm A}^{\rm stat}(x_0,\omega_i)=
-{{1}\over{2}}\left\langle(A_{\rm I}^{\rm stat})_0(x)\, 
{\cal{O}}(\omega_i)\right\rangle \,,
\quad
{\cal O}(\omega)=
{{a^6}\over{L^3}}\sum_{\bf y,z}\overline{\zeta}_{\rm h}({\bf y})\gamma_5
\,\omega({\bf y}-{\bf z})\,\zeta_{\rm l}({\bf z}) \,. 
\label{fastat}
\ee
The boundary-to-boundary correlator $f_1^{\rm stat}$, 
\be
f_1^{\rm stat}(T,\omega_i,\omega_j)=
-{{1}\over{2}}\left\langle{\cal{O}}'(\omega_i)\,
{\cal{O}}(\omega_j)\right\rangle \,,
\quad
{\cal O}'(\omega)=
{{a^6}\over{L^3}}\sum_{\bf y,z}\overline{\zeta}_{\rm l}'({\bf y})\gamma_5
\,\omega({\bf y}-{\bf z})\,\zeta_{\rm h}'({\bf z}) \,,
\label{f1stat}
\ee
serves to cancel the overlap of the (boundary) interpolating fields 
${\cal{O}}(\omega_i)$ with the B-meson state, as made explicit in
\eqs{spect_fastat}~--~(\ref{alphgam}).

As for the choice of $\omega({\bf x})$ itself, we follow \cite{fbstat:pap1} 
and opt for a set of four hydrogen-like, spatially periodic wave functions
\bea
\omega_i({\bf x})
& = &
\frac{1}{N_i}\sum_{\bf n\,\in\,\mathbb{Z}^3}
\overline{\omega}_i\left(|{\bf x-n}L|\right) \,,\qquad 
i=1,\dots,4 \,, 
\nonumber\\[1ex]
\overline{\omega}_1(r)
& = &
r_0^{-3/2}\,\Exp^{\,-r/a_0} \,,\qquad 
\overline{\omega}_2(r)=r_0^{-3/2}\,\Exp^{\,-r/(2a_0)} \,,
\nonumber\\[1ex]
\overline{\omega}_3(r)
& = &
r_0^{-5/2}\,r\,\Exp^{\,-r/(2a_0)} \,,\qquad 
\overline{\omega}_4(\vecx)=L^{-3/2} \,,
\label{wavefcts}
\eea
with $a_0=0.1863r_0$, the hadronic radius $r_0=0.5\,\Fm$ \cite{pot:r0} 
and the normalization factors $N_i$ fixed by
$a^3\sum_{\bf x}\omega_i^2({\bf x})=1$.
Apart from investigating the Schr\"odinger functional correlators for 
single wave functions, it is now also possible (and --- as will become 
clear in the next sections --- even advantageous in practice) to form 
suitable linear combinations of two of them in order to cancel the first 
excited state in the B-meson channel almost completely.

For large $T$ and $x_0$, the correlation functions $\fastat$ and 
$\fonestat$ allow for a computation of the pseudoscalar decay constant in 
the static approximation through the expression for the local 
renormalization group invariant (RGI) matrix element of the static axial 
current,
\be
\Phi_{\rm RGI}^{\rm eff}(x_0,\omega_i)=
-Z_{\rm RGI}\left(1+b_{\rm A}^{\rm stat}am_{\rm q}\right)2L^{3/2}\,
{{f_{\rm A}^{\rm stat}(x_0,\omega_i)}
\over{\sqrt{f_1^{\rm stat}(T,\omega_i,\omega_i)}
}}\,\Exp^{\,(x_0-T/2)E_{\rm eff}(x_0,\omega_i)} \,,
\label{Phi_decay}
\ee
where the effective energy $E_{\rm eff}$ reads 
\be
E_{\rm eff}(x_0,\omega_i)=
\frac{1}{2a}\ln\left[\,
\frac{\fastat(x_0-a,\omega_i)}{\fastat(x_0+a,\omega_i)}\,\right] \,.
\label{Eeff}
\ee
The O($a$) improvement coefficients $\castat$ (cf.~\eq{astatimpr}) and
$\bastat$ depend on the discretization prescription of the static theory and
have been perturbatively determined in~\Ref{HQET:statprec}.
In \eq{Phi_decay}, $\ZRGI$ is the renormalization factor that relates the 
bare matrix elements of $\Astat$ to the RGI ones.
It is non-perturbatively known from~\Ref{zastat:pap3} for the relevant range 
of bare couplings employed here and has a negligible uncertainty in 
comparison to the statistical error associated with the bare matrix element. 
As long as $0\ll x_0\ll T$, the local RGI matrix element is expected to 
exhibit a plateau, from which eventually the value of $\PhiRGI$ to enter the
formula 
\be
\Fps\sqrt{\mps}=
\Cps\left(M/\lMSbar\right)\times\PhiRGI\,+\,\Or\left(1/M\right) 
\label{me_QCD}
\ee
for the pseudoscalar decay constant, $\Fps$, can be extracted.
In this equation, $\mps$ is the meson mass, and the conversion function 
$\Cps$ \cite{zastat:pap3,hqet:pap3} translates the RGI matrix elements of 
the static effective theory to the corresponding QCD matrix elements at 
finite values of the heavy quark mass.
It is parameterized in terms of the RGI mass of the heavy quark ($M$) and 
the QCD $\Lambda$--parameter in the $\MS$ scheme 
($\lMSbar$) \cite{msbar:pap1}.

Before coming to describe our analysis procedure for the computation of
$\PhiRGI$ and the static binding energy based on simulation results for the 
correlation functions in \eqs{fastat} and (\ref{f1stat}), let us have a 
look at the quantum mechanical representation of these correlators.
First, we write down the expressions for $\fastat$ and $\fonestat$ for large 
values of $x_0$ and $T-x_0$, while $L$ remains arbitrary at this stage. 
We neglect terms of order $\exp\{-(T-x_0)\,m_{\rm G}\}$, where the energy 
difference $m_{\rm G}=E^{(0)}_{1}-E^{(0)}_{0}$ is the mass of the $0^{++}$ 
glueball.
In this approximation we obtain the decompositions \cite{msbar:pap2}
\bea
-2\fastat(x_0,\omega_i)
& \approx &
\sum_{k\geq0}\beta_i^{(k)}\rme^{\,-x_0 E_k} \,,\quad
\beta_i^{(k)}=\gamma^{(k)}\alpha^{(k)}_i \,,
\label{spect_fastat}\\[1ex]
2\fonestat(T,\omega_i,\omega_j)
& \approx &
\sum_{k\geq0}\alpha^{(k)}_i\alpha^{(k)}_j\rme^{\,-T E_k} \,.
\label{spect_f1stat}
\eea
Here, the energy $E_0$ of the lowest state can be identified with the 
binding energy $E_{\rm stat}$ of the static-light system, while in addition 
we have introduced
\be                   
\alpha^{(k)}_i=
\frac{\braket{\,k,{\rm PS}\,}{\,{\rm i}_{\,\rm PS}(\omega_i)\,}}
{\braket{\,0,0\,}{\,{\rm i}_{\,0}\,}} \,,\quad
\gamma^{(k)}=
\ketbra{\,0,0\,}{\,\opAstat\,}{\,k,{\rm PS}\,}
\label{alphgam}
\ee
in terms of the $k$-th excited static B-meson state, $|\,k,\rm PS\,\rangle$,
the vacuum $|\,0,0\,\rangle$ and the boundary states \cite{msbar:pap2}
$|\,{\rm i}_{\,\rm PS}(\omega_i)\,\rangle$ and $|\,{\rm i}_{\,0}\,\rangle$, 
all in the finite-volume normalization $\braket{\,\psi\,}{\,\psi\,}=1$.
The desired static-light matrix element, related to the decay constant
according to~\eq{me_QCD}, is then encoded in $\gamma^{(0)}$, because we have
\be
Z_{\rm RGI}\left(1+b_{\rm A}^{\rm stat}am_{\rm q}\right)
\sqrt{2}\,L^{3/2}\times\gamma^{(0)}\equiv
Z_{\rm RGI}\left(1+b_{\rm A}^{\rm stat}am_{\rm q}\right)\Phi_{\rm bare}=
\PhiRGI \,.
\label{gamma}             
\ee
From the above definitions of the correlators one infers that $\PhiRGI$ is 
of mass dimension $3/2$; thus, $r_0^{3/2}\PhiRGI$ is dimensionless. 
\subsection{Observables}
\label{Sec_hlhphys_obs}
According to the foregoing discussion, the Schr\"odinger functional 
correlation functions in \eqs{fastat} and (\ref{f1stat}) obey the following 
asymptotic behaviour for large $x_0$:
\bea
-2f_{\rm A}^{\rm stat}(x_0,\omega_i)
& \,\stackrel{x_0\to\infty}{\sim}\, &
\beta_i^{(0)}\rme^{\,-x_0 E_{\rm stat}}
+\beta_i^{(1)}\rme^{\,-x_0(E_{\rm stat}+\delstat)} \,,
\label{fastat_as}\\[1ex]
2f_1^{\rm stat}(T',\omega_i,\omega_j)
&  \,\stackrel{T'\to\infty}{\sim}\, &
\alpha^{(0)}_i\alpha^{(0)}_j\,\rme^{\,-T'E_{\rm stat}} \,,
\label{f1stat_as}
\eea
where $\delstat$ denotes the energy gap to the first excited state in the
pseudoscalar channel.
In (\ref{f1stat_as}) it is already assumed that, for our values of $T'$, 
contributions to $\fonestat$ from excited states can be neglected 
--- an assumption justified by a numerical analysis in \App{App_f1stat}.
Above, we further exploit the freedom to choose a different time extent 
$T'\neq T$ for the calculation of $\fonestat$; the reason of this choice 
will become clear in the next section.
From the large-time asymptotics of $\fastat$ and $\fonestat$ one concludes
that $\Estat$ and $\PhiRGI$ can be obtained from the associated asymptotic
behaviour of $\Eeff(x_0,\omega_i)$ and $\PhiRGI^{\rm eff}(x_0,\omega_i)$ as
\bea
\hspace{-0.5cm}E_{\rm eff}(x_0,\omega_i)
& \,\stackrel{x_0\to\infty}{\sim}\, &
E_{\rm stat}+\frac{\beta_i^{(1)}}{\beta_i^{(0)}}\,
\frac{\sinh{(a\delstat)}}{a}\,\rme^{\,-x_0\delstat} \,,
\label{Eeff_as}\\[1ex]
\hspace{-0.5cm}\Phi_{\rm RGI}^{\rm eff}(x_0,\omega_i)
& \,\stackrel{x_0\to\infty}{\sim}\, &
\Phi_{\rm RGI}\left\{
1+\frac{\beta_i^{(1)}}{\beta_i^{(0)}}\,\rme^{\,-x_0\delstat}\left[
1+\left(x_0-{\T \frac{T'}{2}}\right)\frac{\sinh(a\delstat)}{a}
\right]\right\} \,.
\label{Phi_as}
\eea
\subsection{Simulation details}
\label{Sec_hlhphys_sim}
%
\TABLE[t]{
\begin{tabular}{ccccccccccccccc}
\toprule
   set                    && $\beta$ && $a\;[\,\Fm\,]$ 
&& $L/a$ && $T/a$ && $T'/a$ && $\kaps$ && $\kapc$   \\
\midrule
   ${\rm A}\,,\,{\rm A}'$ && 6.0     && 0.093      
&& 16    && 24    && 20     && 0.133929 && 0.135196 \\
   ${\rm B}\,,\,{\rm B}'$ && 6.1     && 0.079      
&& 24    && 30    && 24     && 0.134439 && 0.135496 \\
   ${\rm C}\,,\,{\rm C}'$ && 6.2     && 0.068      
&& 24    && 36    && 30     && 0.134832 && 0.135795 \\
   ${\rm D}\,,\,{\rm D}'$ && 6.45    && 0.048      
&& 32    && 48    && 40     && 0.135098 && 0.135701 \\
\bottomrule
\end{tabular}
\caption{%
Simulation parameters for the calculation of the static-light correlation
functions.
Unprimed (primed) data sets refer to volumes $V=L^3\times T$
($V=L^3\times T'$), and the statistics varies between $\Or(1000)$ 
measurements for set ${\rm B}'$ and $\Or(2500-5000)$ measurements for the 
other sets. 
All simulations employed the static action HYP1, except for $\beta=6.45$,
where both versions, HYP1 and HYP2 were used. 
The global periodicity phase in spatial directions of the quark 
fields~\cite{pert:1loop} is set to $\theta=0$ in all cases.
}\label{tab:hqetparam}

}
%
The simulation parameters of our data sets are summarized 
in~\tab{tab:hqetparam}.
The corresponding quenched gauge field ensembles have been generated by a 
standard hybrid overrelaxation algorithm, where each iteration consists of 
one heatbath step followed by a few (in our case five) microcanonical 
reflection steps, and the sequential evaluations of the correlation 
functions were separated by at least $5-10$ iterations.

Thanks to the precise knowledge of the RGI strange quark mass in the
quenched approximation, $\Ms$ \cite{msbar:pap3}, as well as the O($a$)
improved relation between the renormalized PCAC quark mass and the 
subtracted bare quark mass $a\mq=\half\,(\kappa^{-1}-\kapc^{-1})$ for the 
relevant range of bare couplings $\beta=6/g_0^2$ \cite{impr:babp}, the mass 
of the light flavour can be directly fixed to the strange quark mass by 
proper choices of the hopping parameter, $\kappa=\kaps$, without any need 
for interpolations in the light quark mass.
More concretely, our values for $\kaps$ at each $\beta$ were obtained by 
solving $\Ms=\zM\,Z\,(1+\bm\,a\mq)\,\mq$ for $\kappa$, where $\zM$ is known 
from~\cite{msbar:pap1}, $Z$ and $\bm$ from \cite{impr:babp} and the critical 
hopping parameters from \cite{msbar:pap3}.

\section{Analysis strategies for the static-light sector}
\label{Sec_ana}
In this section we describe the extraction of the matrix element of the
static-light axial current and of the static binding energy from the 
correlation functions.
We start with $\Estat$ and display the effective energies for the data sets 
with $\beta=6.45$ and all wave functions in~\fig{fig:Eeff645HYP12}. 

Apart from analyzing the correlation functions and the observables deriving 
from them separately for each wave function, it is advantageous to construct 
linear combinations,
\be
\omega_{ij}= 
{1\over N_{ij}}\left(\omega_i+\rho_{ij}\,\omega_j\right) \,,
\ee
which enhance the quality and the extent of the plateau in the effective 
energy by (approximately) eliminating the contribution of the first excited 
state. 
Since
\be
\rho_{ij}= 
-\beta^{(1)}_i/\beta^{(1)}_j
\ee
achieves this goal exactly, a practical way for finding such linear 
combinations is to perform a simultaneous fit to \eq{fastat_as} for all 
values of $i$, with $\beta^{(0)}_i$, $\beta^{(1)}_i$, $E_{\rm stat}$ and 
$\delstat$ as fit parameters. 
Of course, the fit range has to be chosen with care. 
We list it together with the thus determined $\rho_{ij}$ 
in~\tab{tab:lincomb}.
Errors on the coefficients $\rho_{ij}$ are shown for illustration, but we 
continue the analysis with just the central values.
The so optimized correlation functions are subsequently analyzed without 
assuming anything on which excited state is present.
In other words, we treat them just as if they were arbitrarily chosen trial 
wave functions.
In the above analysis we excluded $\omega_3$, since the time dependence of 
its effective energy, cf.~\fig{fig:Eeff645HYP12}, suggests that several 
states contribute in our chosen range for $x_0$. 
Still, we checked that upon including also $\omega_3$ the final results of 
the next section are unchanged within their uncertainties.
%
\FIGURE[t]{
\epsfig{file=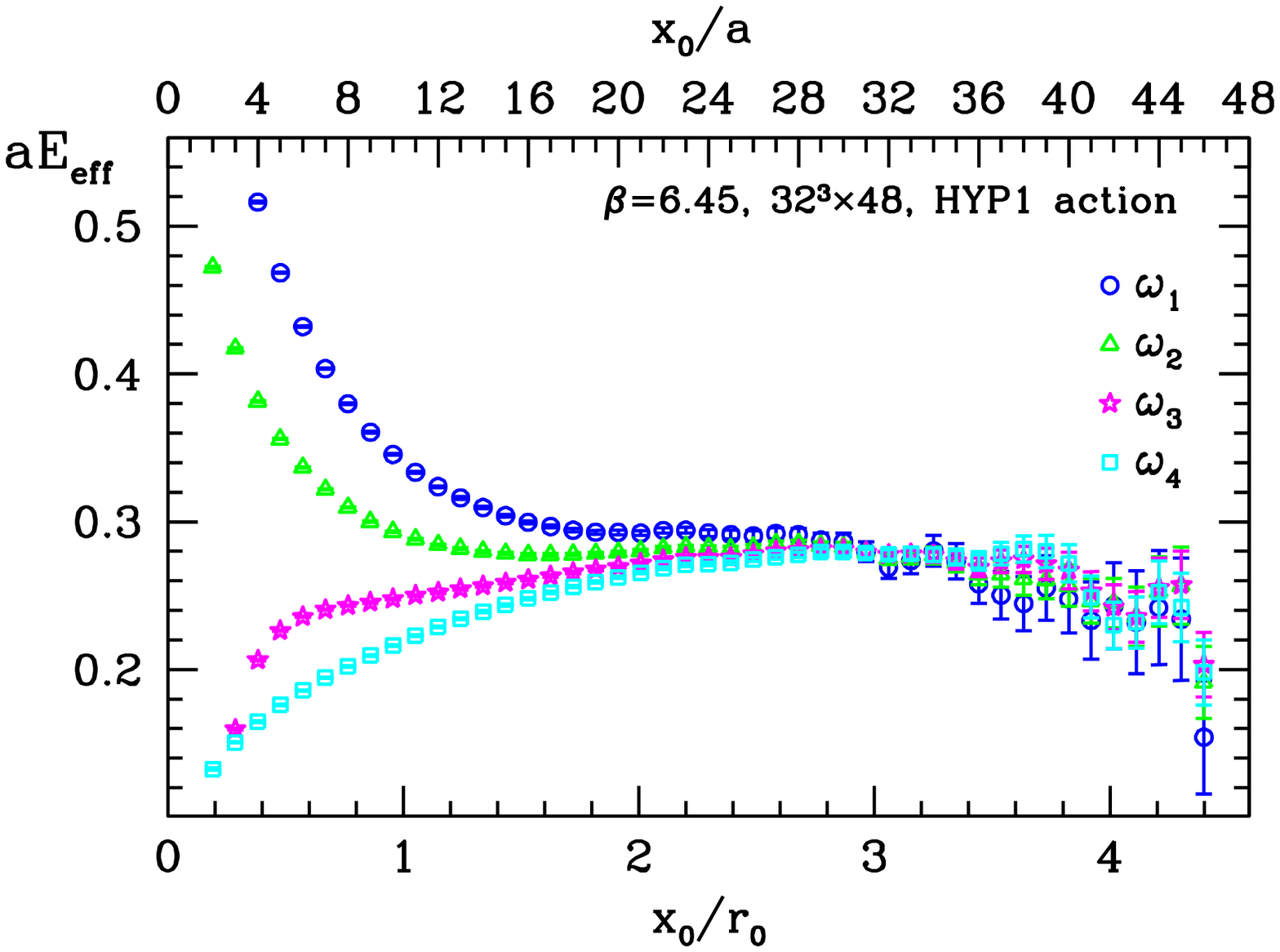,width=0.49\textwidth}
\epsfig{file=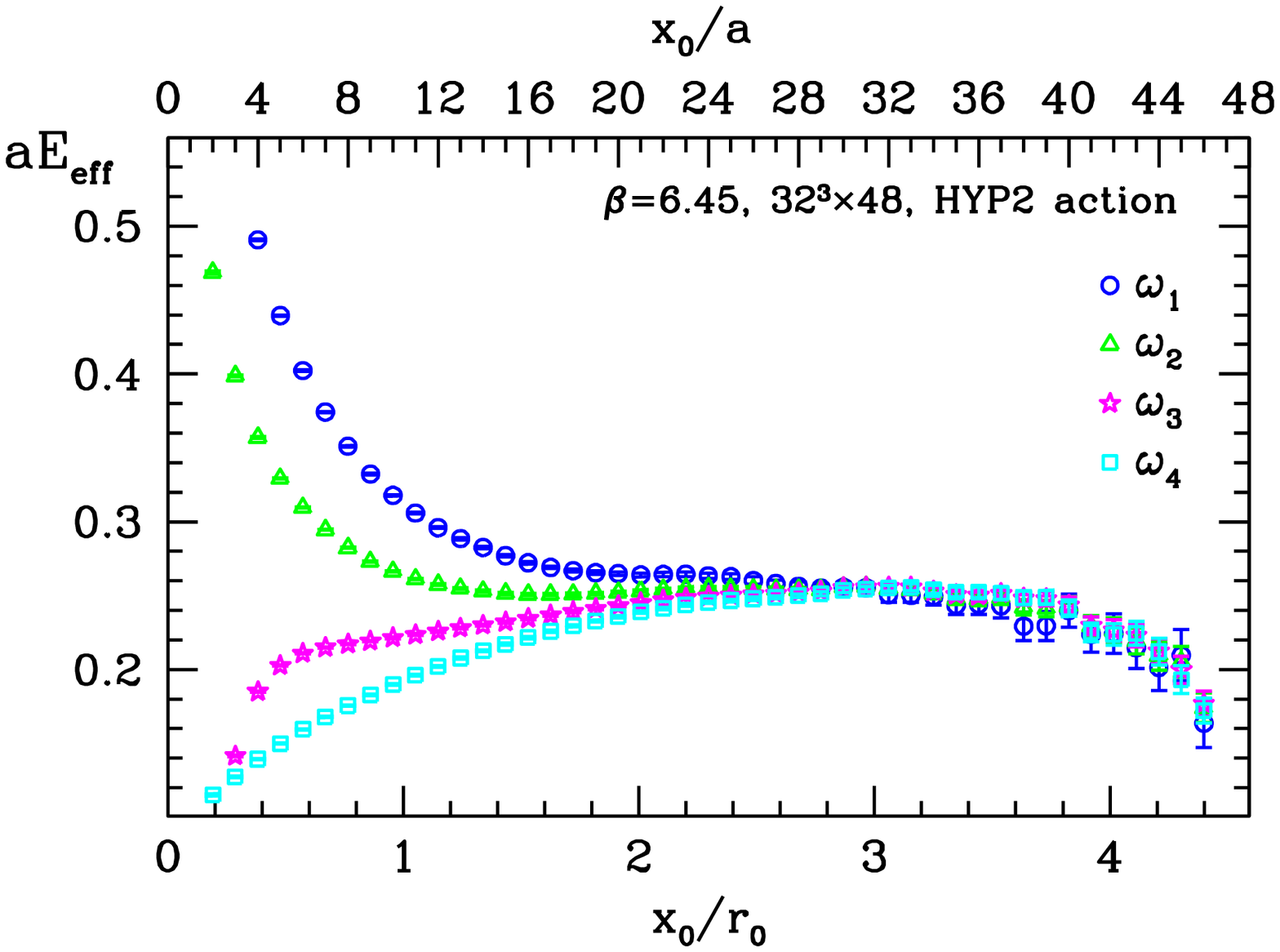,width=0.49\textwidth}
\vspace{-0.5cm}
\caption{%
Effective static-light binding energy from data set D 
($\beta=6.45$, $32^3\times 48$) for the static actions HYP1 (left) and
HYP2 (right).
}\label{fig:Eeff645HYP12}
}
%

After replacing the set of $\omega_i$ with the linear combinations
$\omega_{ij}$, the static binding energy $E_{\rm stat}$ is extracted by fits 
to~\eq{Eeff_as}. 
Either this is done by dropping the correction term and fitting to a 
constant in a rather restricted interval, or we allow for smaller $x_0$ 
where deviations from a plateau are visible and include the correction term. 
These fits again may be performed simultaneously to all linear combinations 
with common fit parameters for the two energies. 
Obviously, having switched to the linear combinations of wave functions, 
the fit parameters in the correction term are expected to refer 
(approximately) to the second excited state. 
Indeed, for instance at $\beta=6.45$, we now obtain a value of 
$a\Delta^{\rm stat}\approx 0.60$, instead of 
$a\Delta^{\rm stat}\approx 0.11$ for the original single wave functions 
with the HYP1 action.
%
\TABLE[t]{
\begin{tabular}{lcccccccr}
\toprule
$\beta$ && $S_{\rm h}$ && $(i,j)$ && $[\,\tmin/a,\tmax/a\,]$ && $\rho_{ij}$ \\
\midrule
6.0  && HYP1 && $(1,2)$ && $[10,18]$ && $1.0(5)$    \\
     && HYP1 && $(1,4)$ && $[10,18]$ && $0.23(5)$   \\
     && HYP1 && $(2,4)$ && $[10,18]$ && $-0.21(5)$  \\
\midrule
6.1  && HYP1 && $(1,2)$ && $[13,20]$ && $\infty$    \\
     && HYP1 && $(1,4)$ && $[13,20]$ && $0.21(9)$   \\
     && HYP1 && $(2,4)$ && $[13,20]$ && $-0.11(10)$ \\
\midrule
6.2  && HYP1 && $(1,2)$ && $[13,24]$ && $1.4(8)$    \\
     && HYP1 && $(1,4)$ && $[13,24]$ && $0.23(6)$   \\
     && HYP1 && $(2,4)$ && $[13,24]$ && $-0.16(6)$  \\
\midrule
6.45 && HYP1 && $(1,2)$ && $[14,30]$ && $1.1(6)$    \\
     && HYP1 && $(1,4)$ && $[14,30]$ && $0.24(6)$   \\
     && HYP1 && $(2,4)$ && $[14,30]$ && $-0.21(6)$  \\
     && HYP2 && $(1,2)$ && $[15,29]$ && $1.5(7)$    \\
     && HYP2 && $(1,4)$ && $[15,29]$ && $0.26(5)$   \\
     && HYP2 && $(2,4)$ && $[15,29]$ && $-0.17(5)$  \\
\bottomrule
\end{tabular}
\caption{%
Coefficients of the linear combinations for all data sets. 
The first linear combination for $\beta=6.1$ coincides with the second wave 
function.
}\label{tab:lincomb}

}
%

As for the effective energy, also for the RGI matrix element of $\Astat$ we 
have investigated several methods of computing this quantity from the 
static-light correlation functions at hand.
For all lattices, and especially at higher $\beta$ (and correspondingly 
larger $T/a$), we observe the dominant part of the statistical uncertainty 
of this quantity to be carried by the correlator between the Schr\"odinger 
functional boundaries, $f_1^{\rm stat}$.
Hence, as already anticipated in~\Sect{Sec_hlhphys_obs}, we have conducted 
additional simulations with temporal extensions $T'<T$, in order to reduce
the error contribution to $\Phi_{\rm RGI}$ from $f_1^{\rm stat}$ by calculating
the latter on lattices with smaller time extents. 
Taking the crucial $\beta=6.45$ data point as an example, 
$f_{\rm A}^{\rm stat}(x_0,\omega_{ij})$ is computed on a $32^3\times T/a$ 
lattice with $T=48a$ and subsequently fitted to a two-state exponential 
ansatz as in \eq{fastat_as}, i.e.
\be
-2f_{\rm A}^{\rm stat}(x_0,\omega_{ij})=
\beta_{ij}^{(0)}\rme^{\,-(x_0-T'/2)E_{\rm stat}}
+\beta_{ij}^{(1)}\rme^{\,-(x_0-T'/2)E_{\rm stat}-x_0\Delta^{\rm stat}} \,,
\label{fastat_2stfit_Tprime}
\ee
whereas $f_1^{\rm stat}$ originates from an \emph{independent} evaluation of
the data set generated in a simulation of a volume of 
$32^3\times T'/a$, $T'=40a$.
The pairs $(T,T')$ for the remaining values of $\beta$ are included
in~\tab{tab:hqetparam}.

In the present situation, the RGI matrix element is given by~\eq{gamma}, 
where $\Phibare$ is reconstructed as
\be
\Phibare(\omega_{ij})=
L^{3/2}\,\frac{\beta_{ij}^{(0)}}
{\sqrt{f_1^{\rm stat}(T',\omega_{ij},\omega_{ij})}} \,. 
\label{Phibare_method3}
\ee
While the non-linear fit parameters $E_{\rm stat}$ and $\Delta^{\rm stat}$ are 
constrained by simultaneous fits of $f_{\rm A}^{\rm stat}$ to 
\eq{fastat_2stfit_Tprime}, the extracted $\Phibare(\omega_{ij})$ are not 
constrained. 
We find them all nicely consistent. 
As an alternative, we also fit  
$\Phibare^{\rm eff}(x_0,\omega_{ij})=\Phibare(\omega_{ij})$ in a restricted 
time interval and found entirely consistent values. 
We quote the latter as our central values.

\section{Results in the static approximation}
\label{Sec_res}
%
\FIGURE[t]{
\epsfig{file=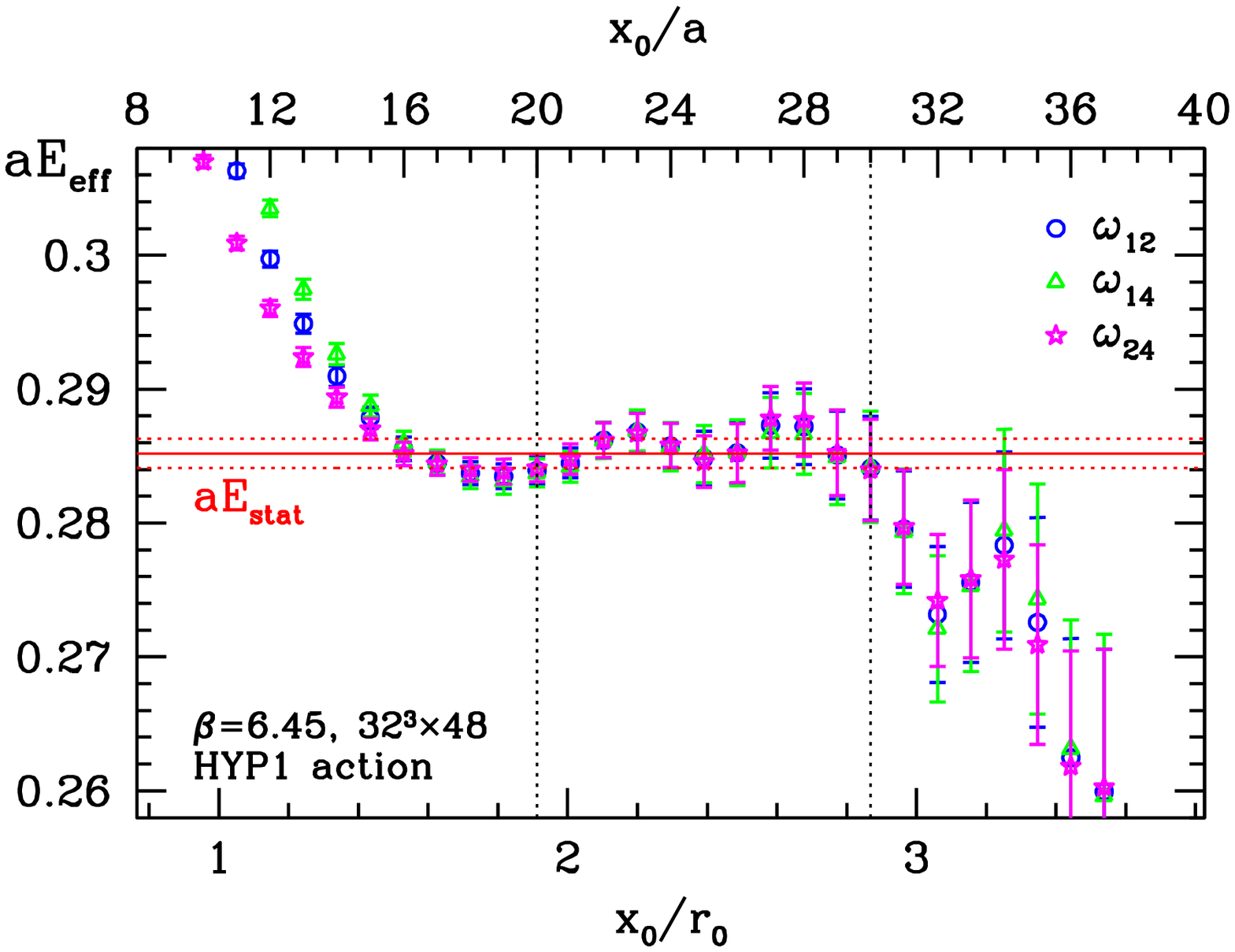,width=0.49\textwidth}
\epsfig{file=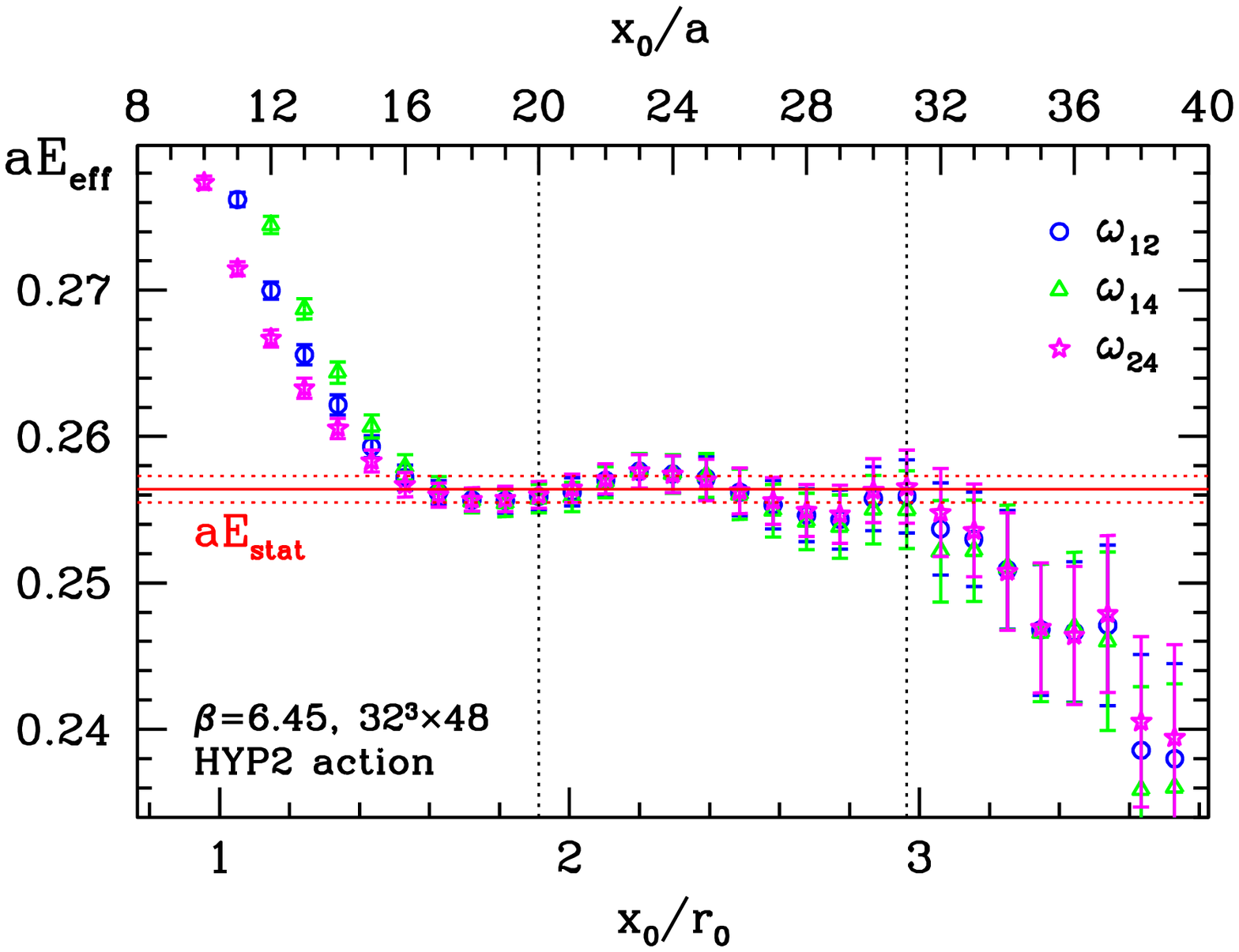,width=0.49\textwidth}
\epsfig{file=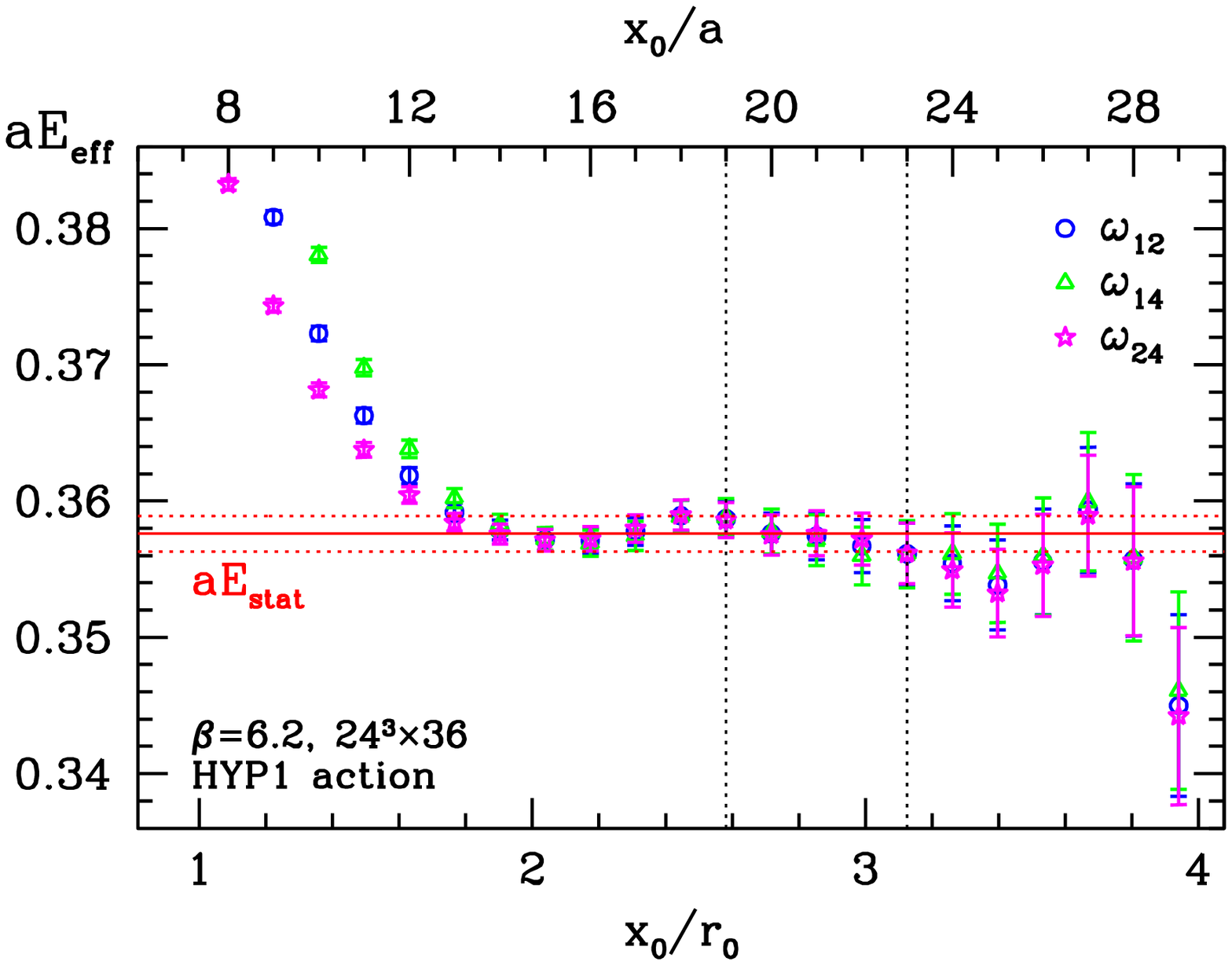,width=0.49\textwidth}
\epsfig{file=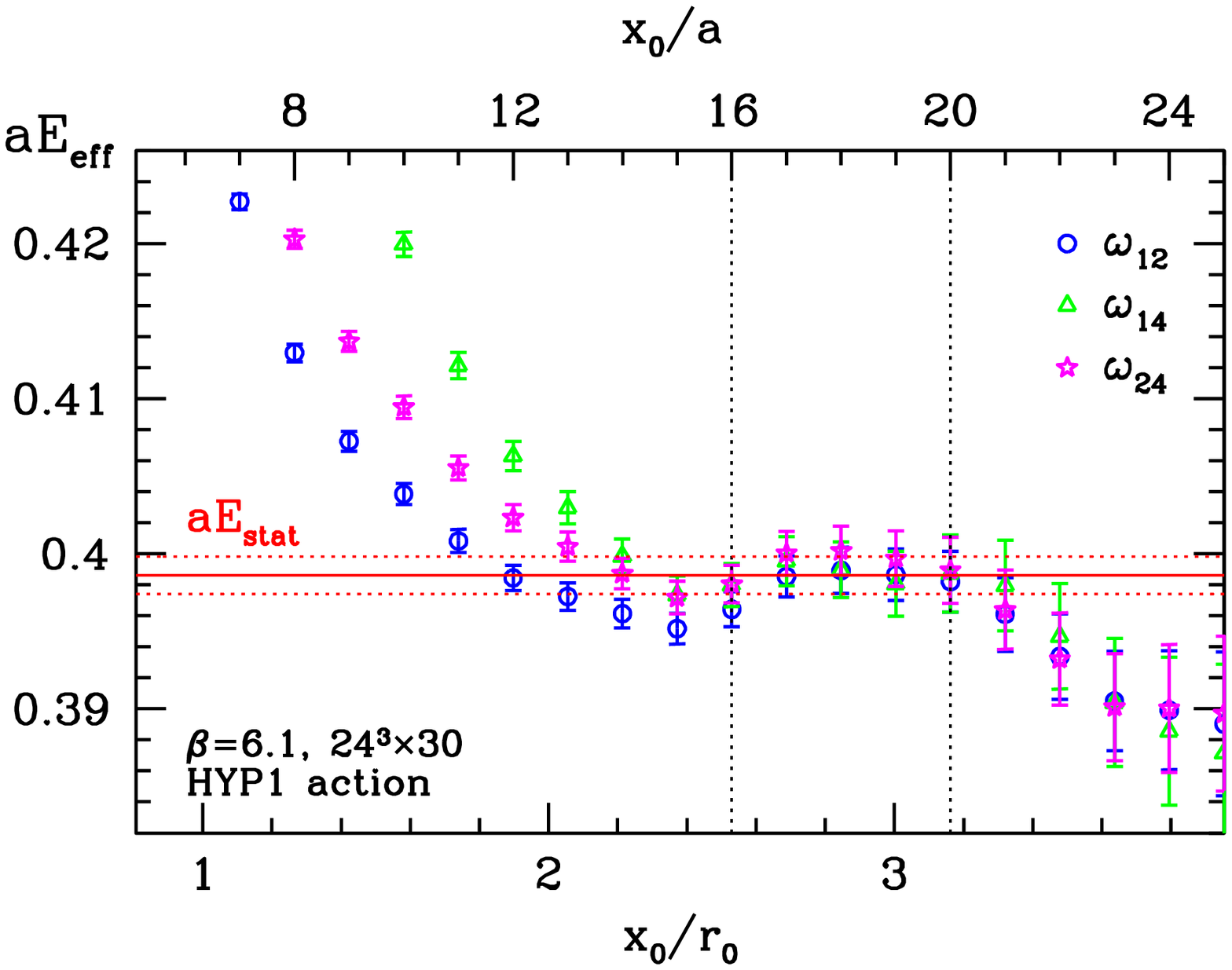,width=0.49\textwidth}
\vspace{-0.5cm}
\caption{%
Top: Effective static-light binding energy from data set D
($\beta=6.45$, $32^3\times 48$) for the static actions HYP1 (left) and HYP2 
(right) after construction of the linear combinations of wave functions, 
labelled $\omega_{12}$, $\omega_{14}$ and $\omega_{24}$ in the panels.  
The final estimates of $\Estat$ (obtained, as explained before, through fits
within intervals given by the vertical dotted lines) are indicated by the 
horizontal lines, where the dashed lines show the error band.
Bottom: Effective binding energies from data sets~C (left) and~B~(right).
}\label{fig:Eeff645HYP12_lincomb}
}
%
We follow the strategy explained in the previous section.
An inspection of the plots for $E_{\rm eff}(x_0,\omega_{ij})$ is very useful 
to get a first impression of the quality and the extension of the plateaux 
as well as to select reasonable fit intervals for the numerical analysis.

Fits are performed in the range $t_{\rm min}\leq x_0\leq t_{\rm max}$, where
$t_{\rm max}$ is suggested by the $x_0$--dependence of 
$E_{\rm eff}(x_0,\omega_{ij})$ and, as already experienced 
in~\Ref{fbstat:pap1}, by the observation that an increase of $t_{\rm max}$ 
beyond a certain threshold (close to $3r_0$) is not convenient, because then 
the statistical uncertainties become too large.
Sensible values for $t_{\rm min}$ may be inferred from the plots as well;
we have kept them such that 
$t_{\rm min}>r_0=0.5\,\Fm$ \cite{pot:r0_ALPHA} holds, while the stability of 
the fit parameters under shifts of 
$t_{\rm min} \to t_{\rm min}-r_0/2$ has always been checked.
In particular, when dropping the excited state contribution, one has to take 
care of the fitting intervals to stay safely inside the plateau region of 
$E_{\rm eff}(x_0,\omega_{ij})$. 
In these cases we usually had $t_{\rm min} \approx 2r_0$. 
Our results are collected in~\tab{tab:res}.
The final error for $\Estat$ is the quadratic sum of the statistical error
and the difference between the values for $\Estat$ obtained by reducing 
$t_{\rm min}$ by $r_0/2$ with $t_{\rm max}$ being fixed.

%
\FIGURE[t]{
\epsfig{file=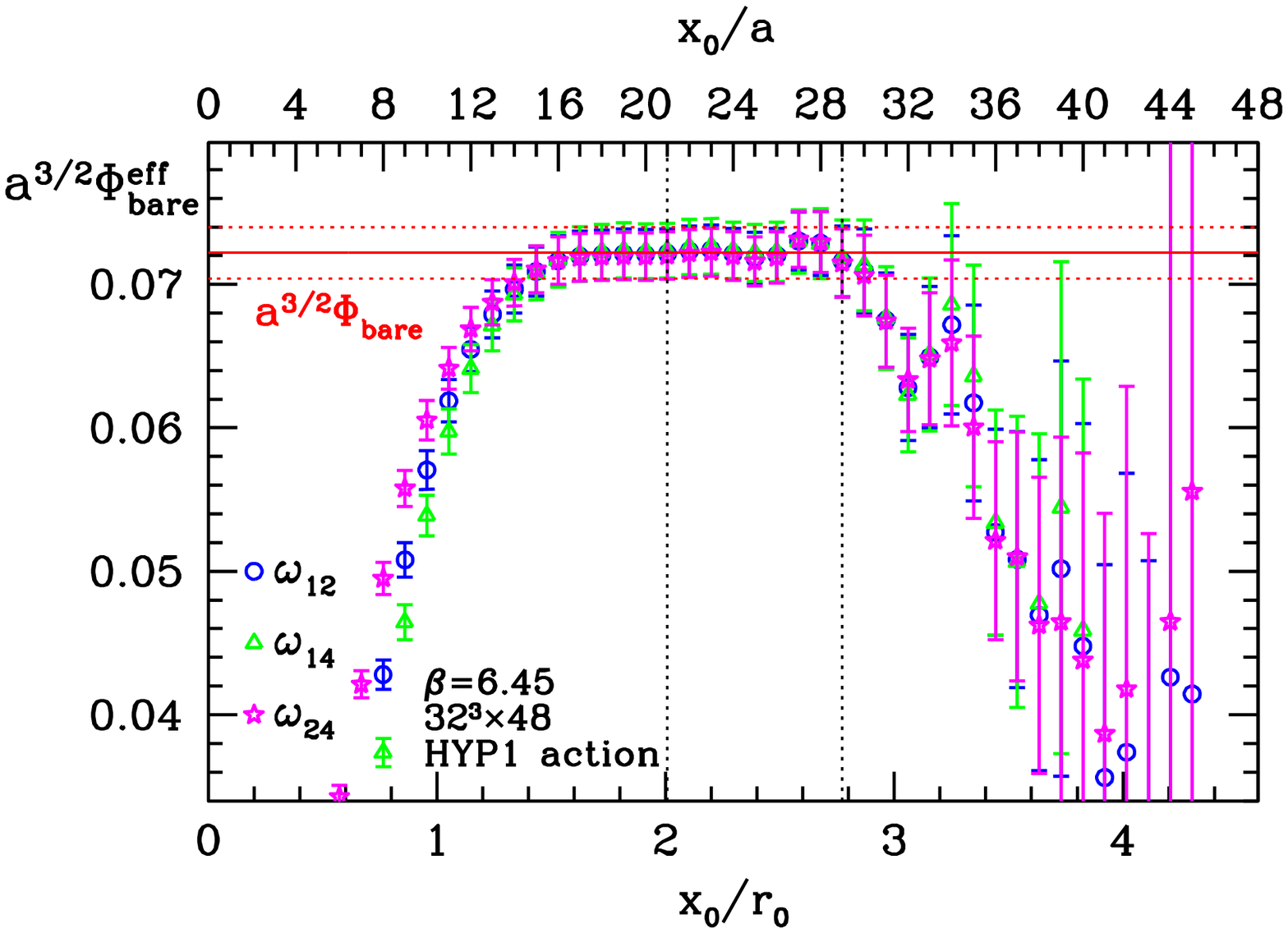,width=0.49\textwidth}
\epsfig{file=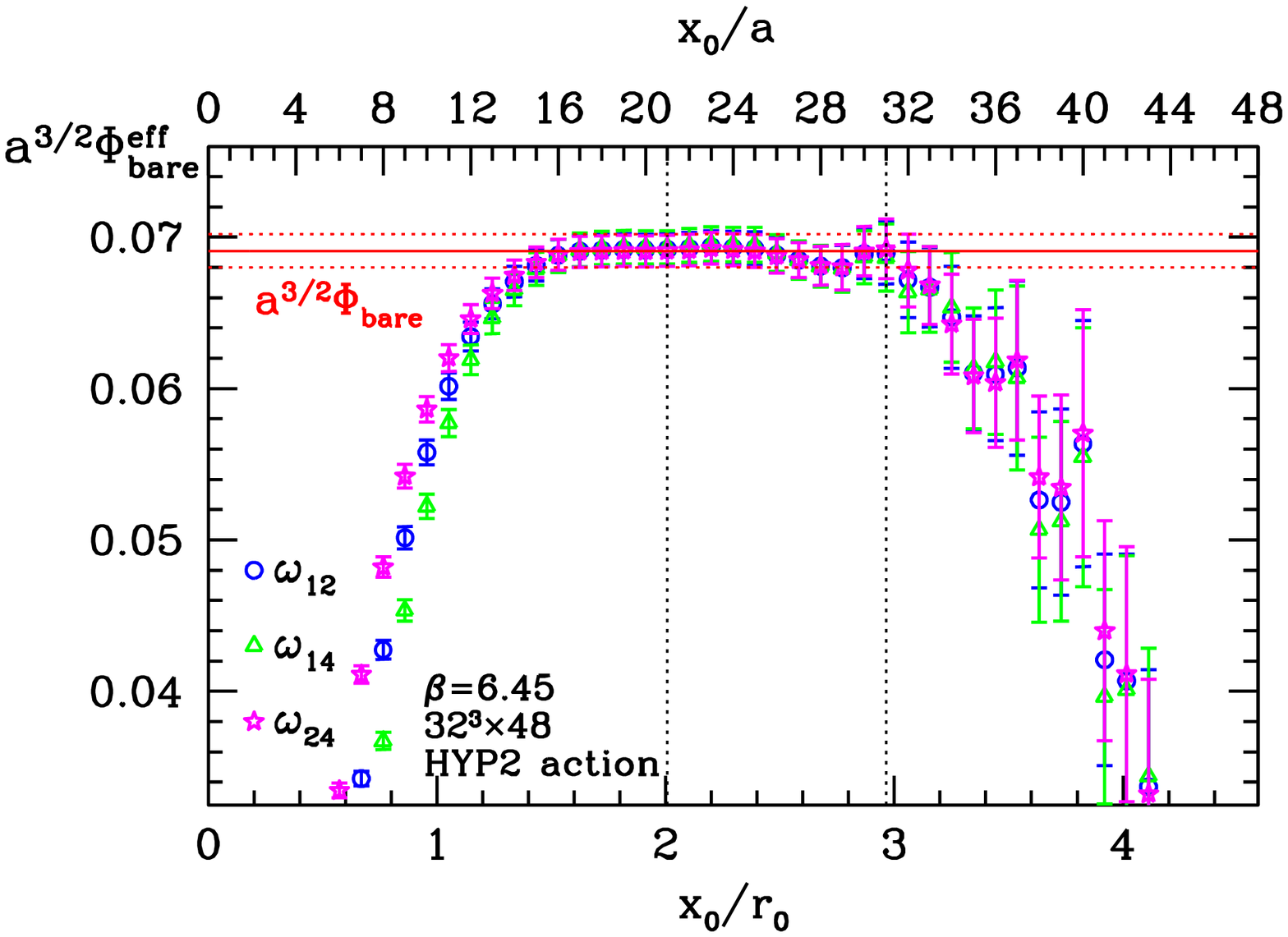,width=0.49\textwidth}
\epsfig{file=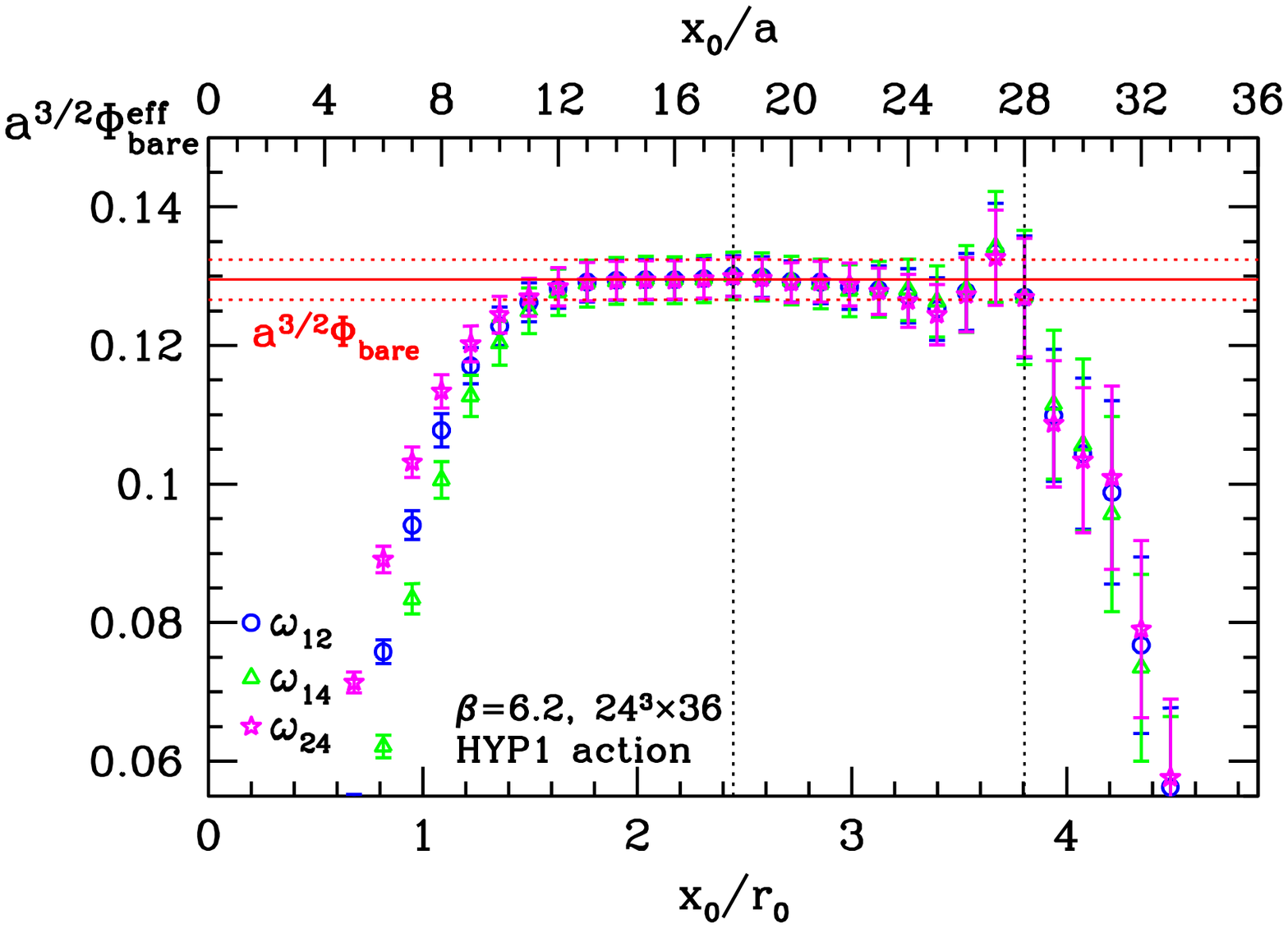,width=0.49\textwidth}
\epsfig{file=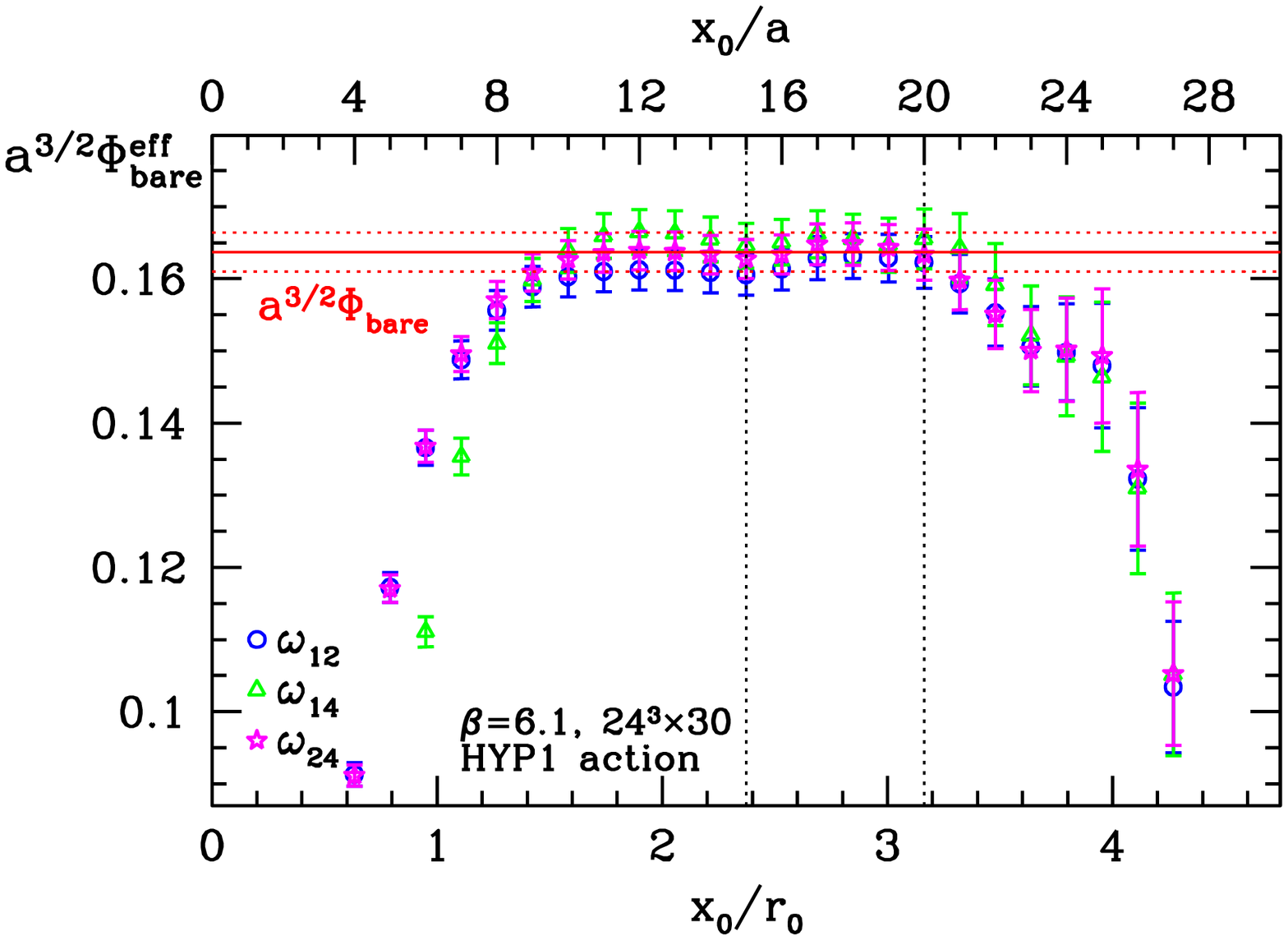,width=0.49\textwidth}
\vspace{-0.5cm}
\caption{%
Top: Local matrix element of the static axial current from data set D
($\beta=6.45$, $32^3\times 48$) for the static actions HYP1 (left) and HYP2 
(right) after construction of the linear combinations of wave functions, 
$\omega_{12}$, $\omega_{14}$ and $\omega_{24}$.
The horizontal lines reflect the result for $a^{3/2}\Phi_{\rm bare}$ and its 
error band (obtained through fits within intervals given by the vertical 
dotted lines).
Bottom: Local matrix elements from data sets C (left) and B (right).
}\label{fig:Phieff645HYP12_lincomb}
}
%
\TABLE[t]{
\begin{tabular}{lcccccc}
\toprule
$\beta$ && $a\Estat$ && $a^{3/2}\Phibare(\omega_{ij})$ && $(i,j)$ \\
\midrule
6.0          && 0.4363(13) && 0.1976(19) && (2,4) \\
6.1          && 0.3986(12) && 0.1637(27) && (2,4) \\
6.2          && 0.3576(13) && 0.1295(29) && (1,2) \\
6.45         && 0.2852(11) && 0.0722(18) && (1,2) \\
$6.45^\star$ && 0.2564(9)  && 0.0691(11) && (1,2) \\ 
\bottomrule
\end{tabular}
\caption{%
Results for the binding energy and the bare static-light decay constant. 
The entries in the last row refer to the HYP2 action.
}\label{tab:res}

}
%
The extraction of the static decay constant from the two different fits
discussed in \Sect{Sec_ana} yields well compatible results.
We list the ones from fits to a constant in~\tab{tab:res} and show them for 
$\beta=6.1-\,6.45$ in~\fig{fig:Phieff645HYP12_lincomb}. 
We checked the dependence of $\PhiRGI$ upon the improvement coefficient 
$\castat$ by setting the latter to its tree-level value and repeating the 
whole analysis. 
The outcome of this exercise is again consistent with the numbers obtained
before.
One can thus conclude that the uncertainty in $\castat$ does not affect our 
results and that they are expected to have all linear $a$--effects removed.
For $\beta\leq 6.2$ and the HYP1 action there is full agreement with the
results already published in~\Ref{fbstat:pap1}. 

With the new data point at $\beta=6.45$ a monotonic $a$--dependence is
observed for $(a/r_0)^2\leq 0.026$. 
Therefore, after attaching the necessary renormalization and improvement
factors (cf.~\eqs{gamma} and (\ref{Phibare_method3})), we perform the 
continuum extrapolation of the RGI matrix element of the static-light axial
current (from the HYP1 data) within this range, 
see~\fig{fig:PhiRGI_contlim}, and within a restricted range
$(a/r_0)^2\leq 0.019$, dropping the data point at $\beta=6.1$ as well.
In the continuum one finds $r_0^{3/2}\PhiRGI=1.624(67)$ and 
$r_0^{3/2}\PhiRGI=1.577(93)$ for the three- and two-point fit, respectively.
As our final estimate we quote the result of the three-point extrapolation 
to ensure stability w.r.t.~statistical fluctuations and supply it with the 
statistical error of the two-point fit as the total uncertainty to try to 
cover a systematic error due to possible higher-order terms in the
$a$--expansion.
This yields
\be
r_0^{3/2}\,\PhiRGI=1.624(93) \,,
\label{PhiRGI_final}
\ee
which agrees within about one standard deviation with our results for both 
static actions at the smallest lattice spacing.
Given the sizeable cutoff effects, it remains desirable to have results at 
an even finer lattice resolution to gain further confidence in the continuum 
extrapolation.
%
\FIGURE[t]{
\epsfig{file=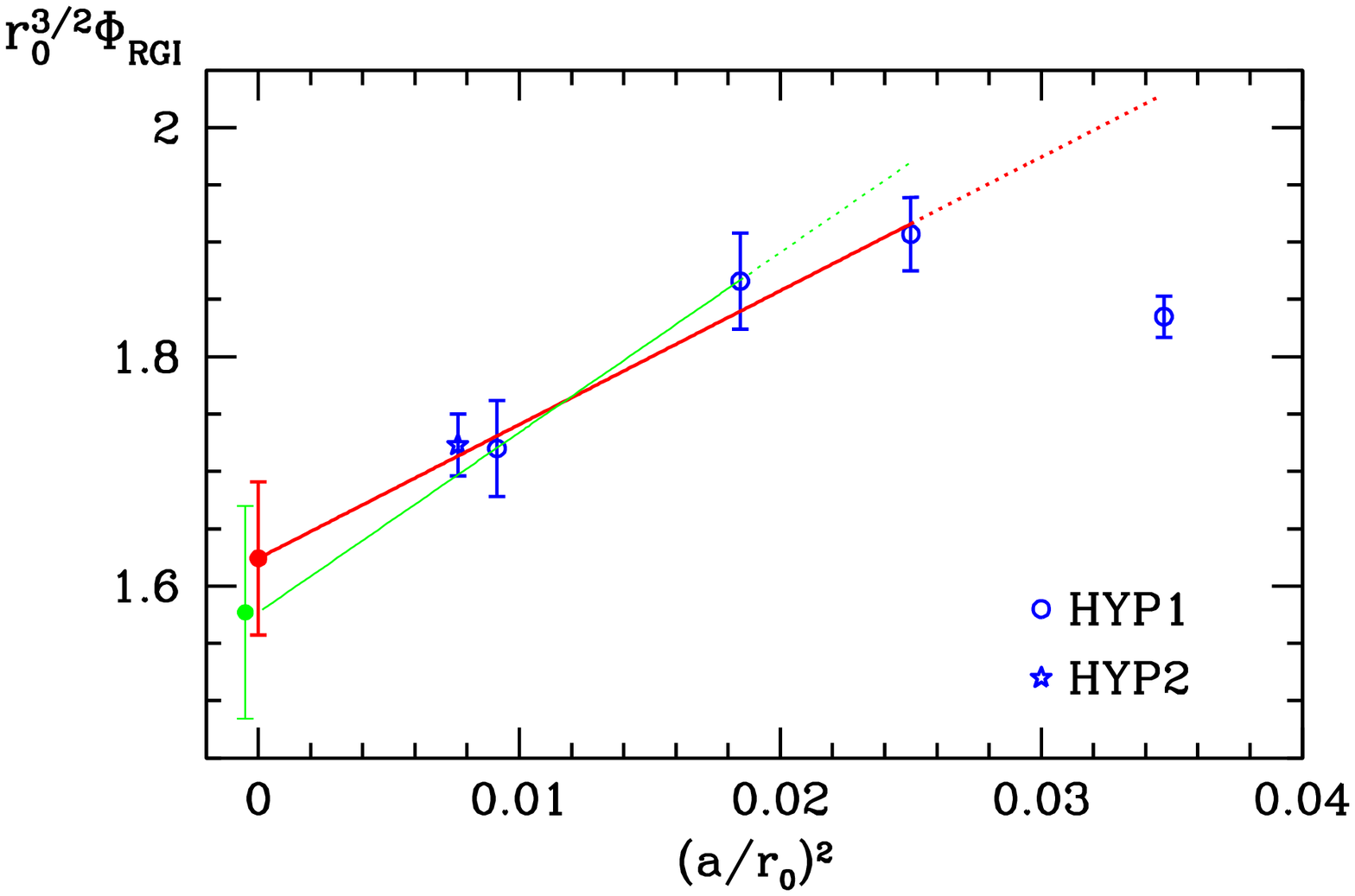,width=0.75\textwidth}
\caption{%
Three- and two-point continuum limit extrapolations of the RGI matrix 
element of the static axial current, represented by the red and the green 
line, respectively.
The HYP2 result at $a\approx 0.05\,\Fm$ is not included in the 
extrapolation but only added for comparison.
The continuum limit from the two-point fit and the HYP2 point are slightly 
moved to the left in the figure.
}\label{fig:PhiRGI_contlim}
}
%

\section{Decay constant at finite heavy quark mass}
\label{Sec_Fbs}
This section details how we compute the pseudoscalar decay constant in the 
continuum limit of large-volume quenched QCD with Schr\"odinger functional 
boundary conditions for pseudoscalar meson masses in the range 
$\mps=(1.5-2.6)\,\GeV$, composed of non-degenerate relativistic quarks.
The heavy quark mass values thus cover a significant range around the 
physical charm region. 
Since it will turn out that the decay constant connects smoothly to the 
static result, $\Phi_{\rm RGI}$, the entire mass region $\mps\geq1.5\,\GeV$ 
is covered and as an application, $\Fbs$ can be extracted. 
\subsection{Relativistic correlation functions and observables}
We compute the decay constant, defined through the QCD matrix element 
between a zero-momentum heavy-light pseudoscalar state and the vacuum, 
from the time component of the ${\rm O}(a)$ improved axial vector current 
on the lattice, 
\be
(A_{\rm I})_{\mu}(x)= 
A_{\mu}(x)+a\,\ca\tilde{\partial}_{\mu}P(x) \,.
\ee
Here, $A_{\mu}(x)$ has the form given in~\eq{astat}, \Sect{Sec_hlhphys_CFs}, 
with the static quark replaced by a heavy relativistic quark, and the 
improvement coefficient $\ca$ in front of the symmetric lattice derivative 
$\tilde\partial_\mu$ acting on the pseudoscalar density $P(x)$ was
non-perturbatively computed in~\cite{impr:pap3}.

$\Or(a)$ improved correlation functions $f_{\rm A}$ and $f_1$ are defined 
just like $f^{\rm stat}_{\rm A}$ and $f^{\rm stat}_1$, but only with the standard
Schr\"odinger functional boundary sources ($\omega=\omega_4={\rm constant}$ 
in \eq{wavefcts}).
Details are as in~\Ref{msbar:pap2}.
For our choice of parameters, the effective mass derived from 
$f_{\rm A}$ exhibits a clear plateau already for these standard Schr\"odinger 
functional boundary sources so that in this part of our calculation we can 
pass on adjusting wave functions to improve the overlap with the ground 
state.
Hence, we readily write down the expressions for the effective heavy-light
pseudoscalar meson mass, 
\be
\mps(x_0)=
\frac{1}{2a}\ln\left[\,
\frac{f_{\rm A}(x_0-a)}{f_{\rm A}(x_0+a)}\,\right] \,, 
\label{meff}
\ee
and the corresponding effective pseudoscalar decay constant
\be
\Fps(x_0)=
-\za\left(1+{\T \frac{\ba}{2}}(am_{{\rm q},i}+am_{{\rm q},{\rm s}})\right)
\frac{2}{\sqrt{\mps L^3}}\,\frac{f_{\rm A}(x_0)}{\sqrt{f_1}}\, 
\Exp^{\,(x_0-T/2)\mps} \,.
\label{Fbs}
\ee
For large enough $T$ and $x_0$ they equal the pseudoscalar mass and decay 
constant. 
Also the renormalization constant $\za$ is non-perturbatively 
known \cite{impr:pap4}, and $am_{{\rm q},i}$ and $am_{{\rm q},{\rm s}}$ are the 
bare subtracted valence quark masses of the heavy and the strange quark, 
respectively.
The coefficient $\ba$, non-perturbatively tuned in~\Ref{impr:losalamos_2},
completes the $\Or(a)$ improvement of our observables. 
\subsection{Simulation details}
%
\TABLE[t]{
\begin{tabular}{ccccccccccc}
\toprule
   $\beta$        && 6.0 && 6.1 && 6.2 && 6.45 && 6.7859 \\
\midrule
   $n_{\rm meas}$ && 380 && 201 && 251 && 289  && 150    \\
\midrule
   $\kappa_1$ && \multicolumn{1}{c}{0.134108} && \multicolumn{1}{c}{0.134548} 
&& \multicolumn{1}{c}{0.134959} && \multicolumn{1}{c}{0.135124} 
&& \multicolumn{1}{c}{0.134739} \\
   $\kappa_2$ && 0.128790 && 0.130750 && 0.131510 && 0.132690 && 0.132440 \\
   $\kappa_3$ && 0.123010 && 0.125870 && 0.127470 && 0.130030 
&& \multicolumn{1}{c}{0.130253} \\
   $\kappa_4$ && \multicolumn{1}{c}{0.119053} && \multicolumn{1}{c}{0.122490}   
&& \multicolumn{1}{c}{0.124637} && \multicolumn{1}{c}{0.128131}   
&& 0.128439 \\
   $\kappa_5$ && 0.115440 && 0.119370 && 0.122000 && 0.126330 && 0.126774 \\
   $\kappa_6$ && 0.112320 && 0.116640 && 0.119680 && 0.124730 && 0.123571 \\
   $\kappa_7$ && 0.109270 && 0.113960 && 0.117370 && 0.123120 && 0.117625 \\
\bottomrule
\end{tabular}
\caption{%
Summary of simulation parameters for the calculation of heavy-light
correlation functions with relativistic quarks.
$\kappa_1\equiv\kaps$ corresponds to the bare subtracted valence quark mass
of the strange quark, $am_{{\rm q},{\rm s}}$, while $\kappa_i$, 
$i=2,\ldots,7$, refer to our choices for the bare subtracted quark mass of
the heavy flavour within the charm region.
}\label{tab:qcdparam}

}
%
We have generated quenched gauge field ensembles for five different lattice 
spacings.
The four coarser lattices have the same $\beta$--values as the ones 
summarized in~\tab{tab:hqetparam}, \Sect{Sec_hlhphys_sim}, and slightly 
different geometries
$(L/a)^3\times T/a=16\times 32$, $24\times 40$, $24\times 48$ and 
$32\times 64$, respectively.
In addition we have generated an ensemble with 
$(L/a)^3\times T/a=48^3\times 96$ and $\beta=6.7859$ which, using 
$r_0=0.5\,\Fm$~\cite{pot:r0}, corresponds to a lattice spacing of 
$a=0.031\,\Fm$~\cite{pot:r0_silvia}.
As in the case of the simulations for the static-light observables, 
we employed a standard hybrid overrelaxation algorithm with 8 to 24 
microcanonical reflection sweeps plus one heatbath sweep forming one 
iteration. 
Subsequent evaluations of the correlation functions were separated by 100 
iterations for $\beta=6.0$, $6.1$, $6.2$ and $6.45$ and by 50 iterations for 
$\beta=6.7859$.
As before, we used the non-perturbatively improved Wilson quark action with 
$\csw$ taken from \cite{impr:pap3}.

In contrast to the hopping parameters in the static-light simulations 
described in \Sect{Sec_hlhphys_sim}, we here have determined $\kaps$ at 
each $\beta$ by fixing the RGI strange quark mass to its value found at 
\emph{finite lattice spacing} in \cite{msbar:pap3}.\footnote{
In order to determine the hopping parameter of the strange quark for the 
lattice with the finest resolution, we were required to extrapolate the 
values for the strange quark mass in~\cite{msbar:pap3} and the quark mass 
renormalization constant to $\beta=6.7859$.
Details can be found in~\cite{thesis:andreasj}.
}
Therefore, $\kappa_1=\kaps$ in~\tab{tab:qcdparam} differs from the choice 
in~\tab{tab:hqetparam} by $\Or(a^3)$. 
The hopping parameter for the physical charm quark is known for the four 
coarser lattices~\cite{mcbar:RS02}, and through an extrapolation it was 
estimated at $\beta=6.7859$. 
We then have guessed further hopping parameters in the vicinity of the charm
quark value such as to yield a homogeneous covering of the region 
$\mps=(1.5-2.6)\,\GeV$ with simulation points.
\subsection{Data analysis and results}
%
\FIGURE[t]{
\epsfig{file=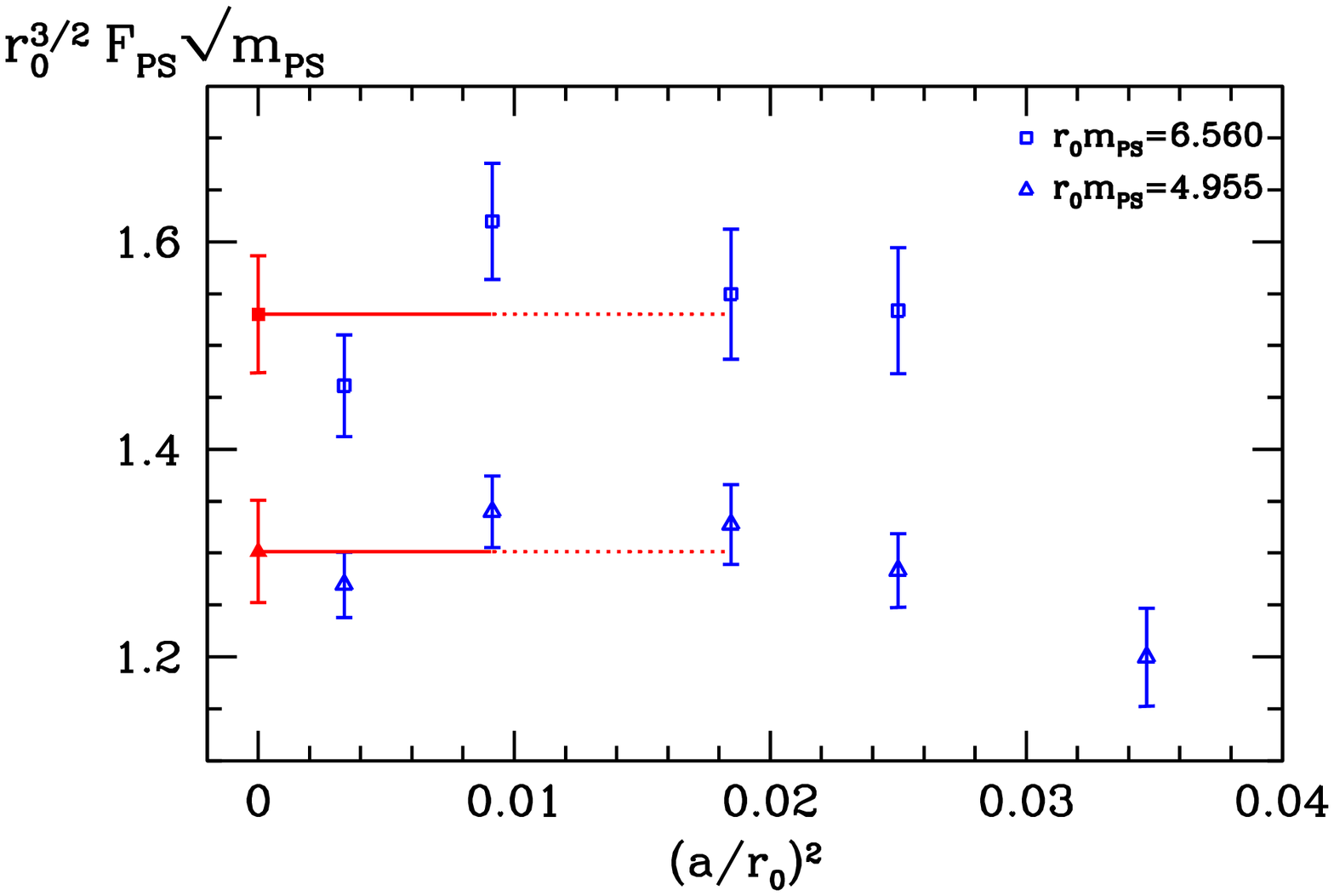,width=0.75\textwidth}
\caption{%
Lattice spacing dependence of the relativistic decay constants for two of 
the larger quark masses.
}\label{fig:fPS_contlim}
}
%
Due to uncertainties in the determination of the simulation parameters,
we do not obtain $F_{\rm PS}\sqrt{m_{\rm PS}}$ for each lattice spacing at 
exactly the same values of $r_0\mps$.  
Therefore, we first interpolate it (at fixed lattice spacing) linearly in 
$1/(r_0\mps)$ to the common points $r_0\mps=3.768$, $4.327$, $4.955$, 
$5.653$, $6.211$ and $6.560$, which are all close to the actual simulation 
points.
In a second step we estimate the continuum limit of the decay constant,
$r_0^{3/2}F_{\rm PS}\sqrt{m_{\rm PS}}$, at fixed $r_0\mps$.

Since we expect $\Or(a^2)$ scaling to break down for too large values of the 
heavy quark masses~\cite{zastat:pap2}, we follow \Ref{HQET:pap2} and exclude 
data points with $aM\gtrsim 0.64$ from the discussion of the continuum 
limit. 
Representative examples for the $a$--dependence are shown 
in~\fig{fig:fPS_contlim}. 
Since the slope of the data at small lattice spacings is not well 
determined, but on the other hand we have results very close to the 
continuum limit itself, we take as our central value for the continuum limit 
the (weighted) average of the data at the two smallest lattice spacings. 
To account for a possible systematic error, we then added the difference 
between the average and the datum at $(a/r_0)^2=0.018$ linearly to the 
statistical uncertainty of this fit to a constant. 

%
\FIGURE[t]{
\epsfig{file=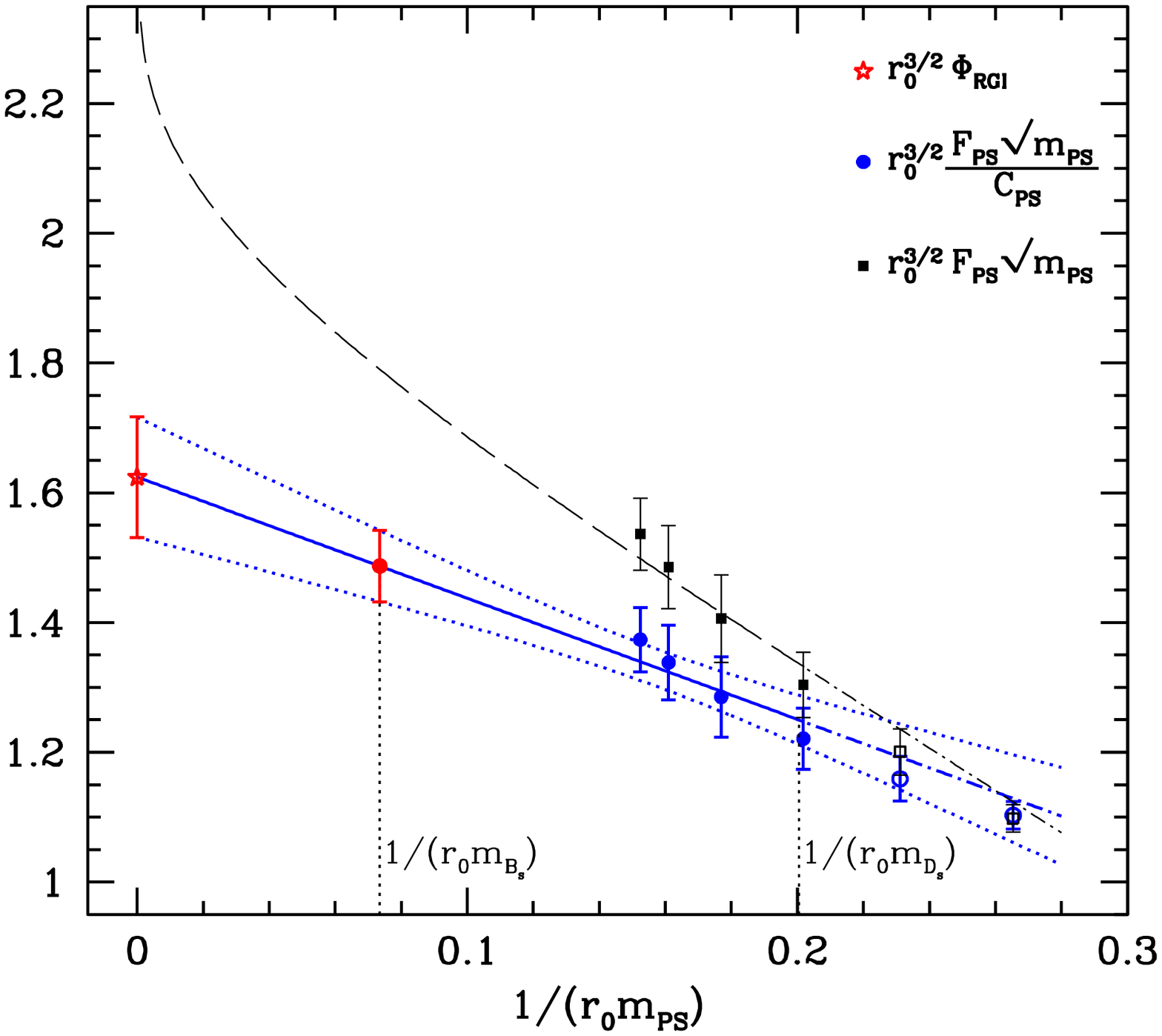,width=0.75\textwidth}
\caption{%
Interpolation of the decay constant between the result in the static limit 
(open red star) and the results in relativistic QCD (blue circles). 
The black squares do not include the matching coefficient $\Cps$. 
Our final quenched result at the physical point of the $\Bs$--meson,
$1/(r_0\mBs)$, is represented by the filled red circle: 
$r_0^{3/2}\Fbs\sqrt{\mBs}/\Cps(\Mb/\Lambda_\msbar)=1.487(55).$
}\label{fig:interpol}
}
%
The outcome of the continuum extrapolations of the relativistic decay
constant for the available six values of the quark mass within the charm
region is illustrated in~\fig{fig:interpol}.
It nicely reflects that our continuum value for the RGI matrix element of 
the static-light axial current, given in~\eq{PhiRGI_final}, can be perfectly
combined with the associated QCD estimates by means of a linear 
interpolation down to even rather low values of the heavy-light meson mass 
of about $1.5\,\GeV$. 

Indeed, as an effective description of the mass dependence of the decay 
constant, we fit the static result and the relativistic data points to the 
form suggested by the HQET expansion,
\be
r_0^{3/2}\,\frac{\Fps\sqrt{\mps}}{\Cps\left(M/\Lambda_\msbar\right)}= 
A\left(1+{B\over r_0\mps}\right) \,,
\label{parameterization}
\ee
where we only include those points among the relativistic data into the fit 
that obey $\mps\gtrsim\mDs$ 
(cf.~the filled blue circles in~\fig{fig:interpol}).
Here, $\Cps$ is the conversion function, which relates the static effective 
theory and QCD and has already appeared in (\ref{me_QCD}).
It is taken from perturbation theory in the form of~\cite{hqet:pap3}; 
its uncertainty is $\Or\left(\alpha(\mps)^3\right)$, estimated to be smaller 
than our statistical errors (see figure 2 in~\cite{hqet:pap3}). 
The resulting slope $B=-1.1(2)$ translates into
\be
B/r_0=-0.45(9)\,\GeV \quad \text{for} \quad r_0=0.5\,\Fm \,.
\label{slope}
\ee
We emphasize that \eq{parameterization} and thus our value for the slope has 
to be considered an effective, phenomenological description and,
consequently, $B$ is an effective slope. 
A true HQET expansion can not be defined without a non-perturbative 
matching of the effective theory and QCD, i.e.~a \emph{non}-perturbative 
definition and estimate of $\Cps$. 
The reason is that at asymptotically large quark mass, any unknown 
perturbative correction in $\Cps$ dominates over the non-perturbative 
$1/m$--term.
We refer to \Ref{reviews:NPRrainer_nara} for a more thorough explanation.

Converting the interpolation result 
$r_0^{3/2}\Fbs\sqrt{\mBs}/\Cps(\Mb/\lMSbar)=1.487(55)$ 
to physical units (using $r_0\lMSbar$ from \cite{msbar:pap1} and
$r_0\Mb$ from \cite{HQET:mb1m} in the evaluation of $\Cps$ as well as the
experimental $\mBs$--value) yields for the quenched decay constant of 
the $\Bs$--meson:
\be
\Fbs=193(7)\,\MeV \,.
\ee
Being obtained from our effective description, this result is, however, 
\emph{not} affected by our previous cautionary remarks on the HQET 
expansion.
All that is needed for its determination is a safe interpolation, which 
\eq{parameterization} does represent.

Finally, note that in an estimate of the $1/\mps$--correction the 
mass dependence in $\Cps$, i.e.~the anomalous dimension of the current in 
the effective theory, plays a numerically important r\^ole. 
This is shown by the squares in \fig{fig:interpol} where, as an 
illustration, $\Cps$ has been dropped. 
The difference reveals that $\Cps$ accounts for about 50\% of the 
mass dependence of $\Fps\sqrt{\mps}$ in the considered region. 

For other calculations of $\Fbs$ we refer 
to~\cite{AliKhan:2001jg,Wingate:2003gm,Onogi:2004gd,fbstat:ukqcd,
Gray:2005ad,Bernard:2007zz,AliKhan:2007tm}
and references therein.

\section{Conclusions}
\label{Sec_concl}
We have presented a computation of the heavy-light pseudoscalar meson decay 
constant in quenched lattice QCD reaching a precision in the continuum limit
of around 4\%. 
The computation is founded on the $\Or(a)$ improvement and non-perturbative 
renormalization of the relativistic theory and the static approximation of 
HQET carried out earlier by the ALPHA Collaboration.
Here, we have added large-volume computations dealing with the heavy quark 
both in the static approximation and in relativistic QCD with masses around
and somewhat heavier than the charm quark's mass.

The b-region is reached through a well controlled interpolation linear in 
the inverse of the meson mass. 
The final result for $\Fbs$ is nicely consistent 
with~\Refs{lat06:damiano,mbfb:Nf0}, where different strategies are applied 
but the same inputs are used to fix the quenched theory.\footnote{
In order to determine the b-quark mass in \cite{HQET:mb1m}, the 
spin-averaged $\Bs$-mass was used instead of the pseudoscalar 
mass~\cite{lat06:damiano,mbfb:Nf0}, but this is a small effect of order 
$\lQCD^3/\mBs^2$.
} 
The \emph{effective} linear slope leads to a 
${\rm O}(1/{m_{\rm PS}})$ correction of about 10\% at the physical b-quark 
mass.
The effective linear pattern is preserved for masses in the charm region, 
and no evidence is found for ${\rm O}(1/{m_{\rm PS}^2})$ corrections at the 
precision level of a few percents.
These conclusions on the mass dependence of the decay constant depend on a 
precise enough knowledge of the conversion function $\Cps$ \cite{hqet:pap3} 
made possible through the perturbative result of \Ref{ChetGrozin}.
It will be very interesting to compare the present result with a direct HQET 
computation including $1/m$--corrections~\cite{lat07:nicolas}.

At the more technical level, an important r\^ole towards a sensitive
reduction of the statistical uncertainty is played by the choice of two 
different physical time extents for the axial current correlator and for the 
boundary-to-boundary correlator in the effective theory, as well as in the 
construction of interpolating fields with wave functions.
By exploiting the latter and the fitting methods used for the decay constant 
we also computed the binding energy of the static-light system for the HYP1 
action in the range of couplings $6.0\leq\beta\leq 6.45$ and at $\beta=6.45$ 
for the HYP2 action. 
They are of interest in a computation of the b-quark 
mass~\cite{HQET:pap1,HQET:mb1m,lat06:damiano,mbfb:Nf0}.
\acknowledgments
This work is part of the ALPHA Collaboration research programme.
We thank NIC/DESY for allocating computer time on the APE computers to this 
project as well as the staff of the computer center at Zeuthen for their 
support.
We further acknowledge partial support by the Deutsche 
Forschungsgemeinschaft (DFG) in the SFB/TR 09-03,
``Computational Particle Physics'', and by the European Community through EU 
Contract No.~MRTN-CT-2006-035482, ``FLAVIAnet''.

\begin{appendix}
\section{Ground state dominance for $\fonestat$}
\label{App_f1stat}
In the description of the analysis method to extract $\PhiRGI$ 
in~\Sect{Sec_ana} we have implicitly assumed that \eq{f1stat_as} holds and 
that contributions from the first and possibly higher excited states to 
$\fonestat$ in \eq{spect_f1stat} can be neglected.

In order to arrive at a quantitative criterion in how far this assumption is 
justified for our data, let us suppress the $\omega$--dependence of the
correlation functions to lighten the notation and write down the quantum 
mechanical representation of $\fonestat$ including the first excited state 
correction, taking over the notation introduced at the end 
of~\Sect{Sec_hlhphys_CFs}:
\be
2f_1^{\rm stat}=
\big[\alpha^{(0)}\big]^2\,\Exp^{\,-T'E_{\rm stat}}\left(
1+\left[\frac{\alpha^{(1)}}{\alpha^{(0)}}\right]^2\,
\Exp^{\,-T'\delstat)}\right) \,.
\ee
By virtue of \eqs{spect_fastat}~--~(\ref{alphgam}), the coefficients 
$\alpha^{(k)}$ are related to the $\beta^{(k)}$ appearing in the 
corresponding decomposition of $\fastat$ through
\be
\left[\frac{\beta^{(1)}}{\beta^{(0)}}\right]^2=
\left[\frac{\alpha^{(1)}}{\alpha^{(0)}}\right]^2\,\left[
\frac{\ketbra{\,0,0\,}{\,\opAstat\,}{\,1,{\rm PS}\,}}
{\ketbra{\,0,0\,}{\,\opAstat\,}{\,0,{\rm PS}\,}}\right]^2=
\left[\frac{\alpha^{(1)}}{\alpha^{(0)}}\right]^2\,
\left[\frac{\Fps^{{\rm stat},(1)}}{\Fps^{{\rm stat}}}\right]^{2} \,,
\ee
where in the second step 
$\ketbra{\,0,0\,}{\,\opAstat\,}{\,k,{\rm PS}\,}\propto
\Fps^{{\rm stat},(k)}\sqrt{\mps}$ 
(with $\Fps^{{\rm stat},(0)}=\Fps^{{\rm stat}}$) has been used.
Since we expect 
\be
\left[\frac{\Fps^{{\rm stat},(1)}}{\Fps^{{\rm stat}}}\right]^{2}=
{\rm O}(1) \,,
\ee
we get 
$\big[\alpha^{(1)}/\alpha^{(0)}\big]^2\sim
\big[\beta^{(1)}/\beta^{(0)}\big]^2$, which leads to\footnote{
Note that $\fonestat$ enters $\PhiRGI$ with a power of $-1/2$,
cf.~\eq{Phi_decay}.
} 
the following correction term to $\PhiRGI$ owing to a second, higher state 
possibly present in $\fonestat$:
\be
\Delta f_1\approx
\frac{1}{2}\left[\frac{\beta^{(1)}}{\beta^{(0)}}\right]^2
\Exp^{\,-T'\delstat} \,.
\label{Delta_f1}
\ee
Here, $\beta^{(0)}$, $\beta^{(1)}$ and $\delstat$ are accessible through the 
two-state fits of $\fastat$ discussed in the main text.

After building linear combinations of the original wave functions, we find 
$\Exp^{\,-T'\delstat}=\Or(10^{-6})$ so that the magnitude of $\Delta f_1$ is 
essentially driven by the ratio of the linear fit parameters $\beta^{(0)}$ 
and $\beta^{(1)}$. 
For all lattices and linear combinations we found $\Delta f_1$ to be orders 
of magnitude smaller than the statistical uncertainty associated with 
$\PhiRGI$ itself, thereby supporting the validity of the one-state dominance 
for $\fonestat$. 
As already pointed out in \Sect{Sec_ana}, thanks to the construction of
linear combinations we actually expect the estimated correction term to be 
predominantly governed by the second excited state instead of the first one 
that is written in the formulae of this appendix.

\end{appendix}
\bibliography{lattice_ALPHA}
\bibliographystyle{JHEP-2}
\end{document}